\documentclass[a4paper,11pt,headings=big,DIV=12]{scrartcl}
\usepackage{multirow}
\usepackage{enumerate}
\usepackage{amsmath,amssymb,marvosym,soul}
\usepackage{bbm}
\usepackage{color}
\usepackage{slashed,fancybox,empheq,accents}
\usepackage[dvipsnames]{xcolor}
\definecolor{darkblue}{rgb}{0,0.2,0.6}
\usepackage[linktoc=page,bookmarks=false,colorlinks=false,
linkbordercolor=RoyalBlue,citebordercolor=ForestGreen,urlbordercolor=CornflowerBlue]{hyperref}
\usepackage{graphicx}
\usepackage{enumerate}
\usepackage[normalem]{ulem}
\usepackage[compress]{cite}
\addtokomafont{disposition}{\rmfamily\boldmath}
\allowdisplaybreaks

\addtokomafont{disposition}{\rmfamily\boldmath}
\let\tmptitle\title\renewcommand{\title}[1]{\tmptitle{\LARGE #1}}
\let\tmpauthor\author\renewcommand{\author}[1]{\tmpauthor{\large #1}}
\let\tmpdate\date\renewcommand{\date}[1]{\tmpdate{\normalsize #1}}

\newcommand{\UtreLC}{$U(3)^3_\mathrm{LC}$}
\newcommand{\UtreRC}{$U(3)^3_\mathrm{RC}$}
\newcommand{\UdueLC}{$U(2)^3_\mathrm{LC}$}
\newcommand{\UdueRC}{$U(2)^3_\mathrm{RC}$}
\newcommand{\Utre}{$U(3)^3$}
\newcommand{\Udue}{$U(2)^3$}

\title{A 125 GeV composite Higgs boson versus flavour and electroweak precision tests}
\author{\normalsize Riccardo Barbieri$^a$, Dario Buttazzo$^a$, Filippo Sala$^a$, David M. Straub$^b$, and Andrea Tesi$^a$
\\{\em\normalsize $^a$Scuola Normale Superiore and INFN, Piazza dei Cavalieri 7, 56126 Pisa, Italy}
\\{\em\normalsize $^b$Institut f\"ur Physik (THEP), Johannes Gutenberg-Universit\"at, 55099 Mainz, Germany}
}
\date{\small{E-mail: \href{mailto:barbieri@sns.it}{barbieri@sns.it}, \href{mailto:dario.buttazzo@sns.it}{dario.buttazzo@sns.it}, \href{mailto:filippo.sala@sns.it}{filippo.sala@sns.it},\\ \href{mailto:david.straub@uni-mainz.de}{david.straub@uni-mainz.de}, \href{mailto:andrea.tesi@sns.it}{andrea.tesi@sns.it}}}

\begin{document}

\maketitle

\begin{abstract}
\noindent
A composite Higgs boson of 125 GeV mass, only mildly fine-tuned, requires top partners with a semi-perturbative coupling and a mass not greater than about a TeV. We analyze the strong constraints on such picture arising from flavour and  electroweak precision tests in models of  {\it partial compositeness}.
We consider different representations for the composite fermions and compare the case of an anarchic flavour structure to models with a $U(3)^3$ and $U(2)^3$ flavour symmetry.
Although non trivially, some models emerge that look  capable of accommodating a 125 GeV Higgs boson  with top partners in an interesting mass range for discovery at the LHC as well as associated flavour signals.
\end{abstract}

\tableofcontents

\section{Introduction}

The discovery of a Higgs-like particle of 125 GeV mass \cite{:2012gk,:2012gu} brings new focus on the longstanding issue of electroweak symmetry breaking (EWSB). Here we are concerned with the implications of this discovery for the view that tries to explain a natural Fermi scale in terms of the Higgs particle as a composite pseudo-Goldstone boson \cite{Kaplan:1983fs,Georgi:1984af,Contino:2003ve, Agashe:2004rs}. More precisely, we shall concentrate our attention on the compatibility of such interpretation of the newly found particle with constraints from flavour and electroweak precision tests (EWPT). 

The common  features  emerging from the modelling of the strong dynamics responsible for the existence of the composite pseudo-Goldstone Higgs boson are: 
\begin{enumerate}[i)]
\item a breaking scale of the global symmetry group, $f$, somewhat larger than the EWSB scale $v \approx 246$~GeV;
\item a set of $\rho$-like vector resonances of typical mass $m_\rho = g_\rho f$;
\item a set of spin-$\frac{1}{2}$  resonances, vector-like under the Standard Model (SM) gauge group, of typical mass $m_\psi = Y f$;
\item bilinear mass-mixing terms between the composite and the elementary fermions, ultimately responsible for the masses of the elementary fermions themselves \cite{Kaplan:1991dc}.
\end{enumerate}

These same mass mixings are crucial in explicitly breaking the global symmetry of the strong dynamics, i.e. in triggering EWSB, with a resulting Higgs boson mass
\begin{equation}\label{mh}
m_h = C \frac{\sqrt{N_c}}{\pi} m_t Y,
\end{equation}
where $N_c=3$ is the number of colors, $m_t$ is the top mass and $C$ is a model dependent coefficient of ${O}(1)$, barring  unnatural fine-tunings \cite{Contino:2006qr, Pomarol:2012qf, Redi:2012ha, Matsedonskyi:2012ym, Marzocca:2012zn, Panico:2012uw}. This very equation makes manifest that the measured mass of 125 GeV calls for a semi-perturbative coupling $Y$ of the fermion resonances and, in turn, for their relative lightness, if one wants to insist on a breaking scale $f$ not too distant from $v$ itself. For a reference value of $f = 500\text{--}700$ GeV,  one expects fermion resonances with typical mass not exceeding about 1 TeV, of crucial importance for their direct searches at the LHC. These searches are currently sensitive to masses in the  500--700 GeV range, depending on the charge of the spin-$\frac{1}{2}$ resonance and on the decay channel~\cite{CMS:2012ab,Chatrchyan:2012vu,Chatrchyan:2012af,ATLAS-CONF-2012-130}.  In this work we aim to investigate the compatibility of this feature with 
flavour and EWPT.

To address this question, we consider a number of  different options for the transformation properties of the spin-$\frac{1}{2}$ resonances under the global symmetries of the strong dynamics, motivated by the need to be consistent with the constraints from the EWPT, as well as different options for the flavour structure/symmetries, motivated by  the many significant flavour bounds. To make the paper readable, after defining the setup for the various cases in section~\ref{sec:setup}, we analyze in succession the different options for the flavour structures/symmetries: {\it Anarchy} in section~\ref{sec:anarchy}, $U(3)^3$ in section~\ref{sec:u3}, $U(2)^3$ in section~\ref{sec:u2}. Section~\ref{sec:ewpt} describes the constraints from EWPT that apply generally to all flavour models. The summary and the conclusions are contained in section~\ref{sec:summary}.

\section{Setup}\label{sec:setup}

To keep the discussion simple and possibly not too model dependent, we follow the {\it partial compositeness} approach of ref.~\cite{Contino:2006nn}. The vector resonances transform in the adjoint representation of a global symmetry respected by the strong sector, which contains the SM gauge group. To protect the $T$ parameter from tree-level contributions, we take this symmetry to be $G_c=SU(3)_c\times SU(2)_L\times SU(2)_R\times U(1)_X$. We assume all vector resonances to have mass $m_\rho$ and coupling $g_\rho$. For the explicit form of their effective Lagrangian we refer to \cite{Contino:2006nn}.

The choice of the fermion representations has important implications for the electroweak precision constraints. We will consider three cases, as customary in the literature.
\begin{enumerate}
\item The elementary $SU(2)_L$ quark doublets, $q_L$, mix with composite vector-like $SU(2)_L$ doublets, $Q$, one per generation. The elementary quark singlets, $u_R$ and $d_R$, couple both to an $SU(2)_R$ doublet $R$. We will call this the {\bf doublet model}.
\item The elementary $SU(2)_L$ quark doublets mix with a  composite $L = (2,2)_{2/3}$ of $SU(2)_L\times SU(2)_R\times U(1)_X$, and the elementary quark singlets couple both to a composite triplet $R=(1,3)_{2/3}$. The model also contains a $(3,1)_{2/3}$ to preserve LR symmetry.  We will call this the {\bf triplet model}.
\item The elementary $SU(2)_L$ quark doublets mix with a $L_U = (2,2)_{2/3}$ and a $L_D = (2,2)_{-1/3}$ of $SU(2)_L\times SU(2)_R\times U(1)_X$, the former giving masses to up-type quarks, the latter to down-type quarks. The elementary up and down quark singlets couple to a $(1,1)_{2/3}$ and a $(1,1)_{-1/3}$ respectively. We will call this the {\bf bidoublet model}.
\end{enumerate}

For concreteness, the part of the Lagrangian involving fermions  reads
\begin{itemize}
\item In  the doublet model
\begin{gather}\label{doubletL}
\mathcal L_s^\text{doublet} =\
-\bar Q^i m_{Q}^i Q^i
-\bar R^i m_{R}^i R^i
+ \left( Y^{ij} \text{tr}[ \bar Q^i_L \mathcal H R_R^j]  + \text{h.c} \right)
\,,
\\
\mathcal L_\text{mix}^\text{doublet} =
m_{Q}^j\lambda_{L}^{ij}\bar q_L^i Q_{R}^j
+
m_{R}^i\lambda_{Ru}^{ij}\bar U_L^i u_{R}^j
+
m_{R}^i\lambda_{Rd}^{ij}\bar D_L^i d_{R}^j
\,.
\end{gather}
where $\mathcal H=(i\sigma_2H^*,H)$ and $R=(U~D)^T$ is an $SU(2)_R$ doublet;

\item  In the triplet model
\begin{gather}
\mathcal L_s^\text{triplet} =\
-\text{tr}[ \bar L^i m_{L}^i L^i ]
-\text{tr}[ \bar R^i m_{R}^i R^i ]
-\text{tr}[ \bar R^{\prime\, i} m_{R}^iR^{\prime\, i}]\notag\\
\qquad\qquad+ Y^{ij} \text{tr}[ \bar L_L^i \mathcal H R_R^j] + Y^{ij}\text{tr}[\mathcal H\,  \bar L_L^i R_R^{\prime\, j}]  + \text{h.c}
\,,
\label{tripletL} \\
\mathcal L_\text{mix}^\text{triplet} =
m_{L}^j\lambda_{L}^{ij}\bar q_L^i Q_{R}^j
+
m_{R}^i\lambda_{Ru}^{ij}\bar U_L^i u_{R}^j
+
m_{R}^i\lambda_{Rd}^{ij}\bar D_L^i d_{R}^j
\,.
\end{gather}
where $Q$ is the $T_{3R}=-\frac{1}{2}$ $SU(2)_L$ doublet contained in $L$ and $U, D$ are  the elements in the triplet $R$ with charge 2/3 and -1/3 respectively;
\item In the bidoublet model
\begin{gather}\label{bidoubletL}
\mathcal L_s^\text{bidoublet} =\
-\text{tr}[ \bar L_U^i m_{Q_u}^i L_U^i ]
-\bar U^i m_{U}^i U^i
+ \left( Y_U^{ij} \text{tr}[ \bar L_U^i \mathcal H ]_L U_R^j  + \text{h.c} \right)
+ (U\to D)
\,,
\\
\mathcal L_\text{mix}^\text{bidoublet} =
m_{Q_u}^j\lambda_{Lu}^{ij}\bar q_L^i Q_{Ru}^j
+
m_{U}^i\lambda_{Ru}^{ij}\bar U_L^i u_{R}^j
+ (U,u\to D,d)
\,,
\end{gather}
where again $Q_u$ and $Q_d$ are the doublets in $L_U$ and $L_D$ which have the same gauge quantum numbers of the SM left-handed quark doublet.
\end{itemize}
Everywhere $i,j$ are flavour indices.
The field content in all three cases is summarized in table~\ref{tab:fields}.\footnote{Note that we have omitted ``wrong-chirality'' Yukawa couplings like $\tilde Y^{ij}{\rm tr}[\bar Q_R^i\mathcal{H}R_L^j]$ for simplicity. They are not relevant for the tree-level electroweak and flavour constraints and do not add qualitatively new effects to the loop contributions to the $T$ parameter.}

We avoid an explicit discussion of the relation between the above simple effective Lagrangians and more basic models which include the Higgs particle as a pseudo-Goldstone boson. Here it suffices to say that the above Lagrangians are suitable to catch the main phenomenological properties of more fundamental models. For this to be the case, the truly basic assumption is that the lowest elements of towers of resonances, either of spin-$\frac{1}{2}$ or of spin 1, normally occurring in more complete models, are enough to describe the main phenomenological consequences, at least in as much as tree-level effects are considered.
For simplicity we also assume the composite fermions to have all the same mass.
To set the correspondence between the {\it partial compositeness} Lagrangians that we use and models with the Higgs as a pseudo-Goldstone boson, one can take the composite Yukawa couplings $Y^{ij}$ in \eqref{doubletL},\eqref{tripletL} and \eqref{bidoubletL} to be proportional to the parameter $Y$ in \eqref{mh}, and identify the common fermion mass with $m_\psi =Yf$, up to a model dependent factor of ${O}(1)$.

\begin{table}
\renewcommand{\arraystretch}{1.15}
\centering
\begin{tabular}{lccccc}
\hline
model && $SU(3)_c$ & $SU(2)_L$ & $SU(2)_R$ & $U(1)_X$ \\
\hline
\multirow{2}{1.5cm}{doublet}
&$Q$ & $\mathbf{3}$ & $\mathbf{2}$ & $\mathbf{1}$ & $\frac{1}{6}$ \\
&$R$ & $\mathbf{3}$ & $\mathbf{1}$ & $\mathbf{2}$ & $\frac{1}{6}$ \\
\hline
\multirow{3}{1.5cm}{triplet}
&$L$ & $\mathbf{3}$ & $\mathbf{2}$ & $\mathbf{2}$ & $\frac{2}{3}$ \\
&$R$ & $\mathbf{3}$ & $\mathbf{1}$ & $\mathbf{3}$ & $\frac{2}{3}$ \\
&$R'$ & $\mathbf{3}$ & $\mathbf{3}$ & $\mathbf{1}$ & $\frac{2}{3}$ \\
\hline
\multirow{4}{1.5cm}{bidoublet}
&$L_U$ & $\mathbf{3}$ & $\mathbf{2}$ & $\mathbf{2}$ & $\frac{2}{3}$ \\
&$L_D$ & $\mathbf{3}$ & $\mathbf{2}$ & $\mathbf{2}$ & $-\frac{1}{3}$ \\
&$U$ & $\mathbf{3}$ & $\mathbf{1}$ & $\mathbf{1}$ & $\frac{2}{3}$ \\
&$D$ & $\mathbf{3}$ & $\mathbf{1}$ & $\mathbf{1}$ & $-\frac{1}{3}$ \\
\hline
\end{tabular}
\caption{Quantum numbers of the fermionic resonances in the three models considered. All composite fields come in vector-like pairs. The $X$ charge is related to the standard hypercharge as $Y=T_{3R}+X$.}
\label{tab:fields}
\end{table}

\subsection{Flavour structure}\label{sec:flasy}

Quark masses and mixings are generated after electroweak symmetry breaking from the composite-elementary mixing.  The states with vanishing mass at $v=0$ obtain the standard Yukawa couplings, in matrix notation,
\begin{align}\label{SMYuk}
\hat y_u &\approx s_{Lu} \cdot U_{Lu} \cdot Y_U \cdot U_{Ru}^\dagger \cdot s_{Ru}
\end{align}
where
\begin{gather}
\lambda_{Lu}=\text{diag}(\lambda_{Lu1},\lambda_{Lu2},\lambda_{Lu3}) \cdot U_{Lu}\,,
\\
\lambda_{Ru}=U_{Ru}^{\dag}\cdot \text{diag}(\lambda_{Ru1},\lambda_{Ru2},\lambda_{Ru3})\,,
\\
s_{X}^{ii}={\lambda_{X i}}/{\sqrt{1+(\lambda_{X i})^2}},
~ X = L,R~,
\end{gather}
and similarly for $\hat y_d$. Here and in the following, the left-handed mixings are different for $u$ and $d$ quarks, $s_{Lu}\neq s_{Ld}$, only in the bidoublet model.
At the same time, in the $v=0$ limit, the remaining states have mass $m_{\psi}$ or $m_{\psi}/\sqrt{1+(\lambda_X)^2}$, respectively if they mix or do not mix with the elementary fermions.

While the effective Yukawa couplings $\hat y_{u,d}$ must have the known hierarchical form, the Yukawa couplings in the strong sector, $Y_{U,D}$, could be structureless {\em anarchic} matrices (see e.g. \cite{Grossman:1999ra,Huber:2000ie,Gherghetta:2000qt,Agashe:2004cp,Blanke:2008zb,Bauer:2009cf,KerenZur:2012fr,Csaki:2008zd}).
However, to ameliorate flavour problems, one can also impose global flavour symmetries on the strong sector. We discuss three cases in the following.

\subsubsection*{Anarchy}

In the anarchic model, the $Y_{U,D}$ are anarchic matrices, with all entries of similar order, and the Yukawa hierarchies are generated by hierarchical mixings $\lambda$.
From a low energy effective theory point of view, the
requirement to reproduce the observed quark masses and mixings fixes the relative size of the mixing parameters up to -- a priori unknown -- functions of the elements in $Y_{U,D}$. We follow the common approach to replace functions of Yukawa couplings by appropriate powers of ``average'' Yukawas $Y_{U*,D*}$, keeping in mind that this introduces $O(1)$ uncertainties in all observables. In this convention, assuming $\lambda_{X3}\gg\lambda_{X2}\gg\lambda_{X1}$, the quark Yukawas are given by
\begin{align}
y_{u}&=Y_{U*} s_{Lu 1}s_{Ru 1}~,&
y_{c}&=Y_{U*} s_{Lu 2}s_{Ru 2}~,&
y_{t}&=Y_{U*} s_{Lu 3}s_{Ru 3}~.
\label{eq:anarchy-yukawas}
\end{align}
and similarly for the $Q=-1/3$ quarks. In the doublet and triplet models, the entries of the CKM matrix are approximately given by
\begin{align}
V_{ij} &\sim \frac{s_{L i}}{s_{L j}} ~,
\label{eq:anarchy-ckm-doublet}
\end{align}
where $i<j$. Using eqs.~(\ref{eq:anarchy-yukawas}) and (\ref{eq:anarchy-ckm-doublet}), one can trade all but one of the $s_{L,R}$ for known quark masses and mixings. We choose the free parameter as
\begin{equation}
x_t \equiv s_{L3}/s_{Ru3}.
\label{eq:x-doublet}
\end{equation}
In the bidoublet model, instead of (\ref{eq:anarchy-ckm-doublet}) one has
 in general two different contributions to $V_{ij}$%
\begin{align}
V_{ij} &\sim
\frac{s_{Ld i}}{s_{Ld j}}\pm \frac{s_{Lu i}}{s_{Lu j}} ~.
\label{eq:anarchy-ckm-bidoublet}
\end{align}
Given the values of all quark masses and mixings, the hierarchy $\lambda_{X3}\gg\lambda_{X2}\gg\lambda_{X1}$ is only compatible with $s_{Lu i}/s_{Lu j}$ being at most comparable to $s_{Ld i}/s_{Ld j}$. In view of this, the two important parameters are
\begin{align}
x_t &\equiv s_{Lt}/s_{Rt} ~,
&
z &\equiv s_{Lt}/s_{Lb} ~.
\label{eq:x-bidoublet}
\end{align}
The requirement to reproduce the large top quark Yukawa ($m_t = \frac{y_t}{\sqrt{2}}v$)
\begin{equation}
y_t = s_{Lt} Y_{U*} s_{Rt},
\end{equation}
restricts $x_t$ to a limited range around one\footnote{\text{In our numerical analysis, we will take $y_t=0.78$, which is the running $\overline{\text{MS}}$ coupling at 3~TeV.}},
\begin{equation}
\frac{y_t}{Y_{U*}} < x_t < \frac{Y_{U*}}{y_t} ~,
\end{equation}
while we take $z$ throughout to be greater than or equal to 1.

From now on we identify $Y_{U*}$ and $Y_{D*}$ with the parameter $Y$ of \eqref{mh}.

\subsubsection*{$U(3)^3$}

In the $U(3)^3$ models \cite{Cacciapaglia:2007fw,Barbieri:2008zt,Redi:2011zi} one tries to ameliorate the flavour problem of the anarchic model by imposing a global flavour symmetry, at the price of giving up the potential explanation of the generation of flavour hierarchies. Concretely,
one assumes the strong sector to be invariant under the diagonal group $U(3)_{Q+U+D}$ or $U(3)_{Q^u+U}\times U(3)_{Q^d+D}$. The composite-elementary mixings are the only sources of breaking of the flavour symmetry of the composite sector and of the $U(3)_{q}\times U(3)_{u}\times U(3)_{d}$ flavour symmetry of the elementary sector. We consider two choices.
\begin{enumerate}
\item In {\it left-compositeness}, to be called {\boldmath\UtreLC} in short, the left mixings are proportional to the identity, thus linking $q$ to $Q~(Q^u, Q^d)$ into $U(3)_{Q+U+D+q}$ (or $U(3)_{Q^u+Q^d+U+D+q}$),
and the right mixings $\lambda_{Ru}$, $\lambda_{Rd}$ are the only source of $U(3)^3$ breaking.

\item In {\it right-compositeness}, to be called {\boldmath\UtreRC} in short, the right mixings link $u$ to $U$ and $d$ to $D$ into $U(3)_{Q^u+U+u}\times U(3)_{Q^d+D+d}$, while
the left mixings $\lambda_{Lu}$, $\lambda_{Ld}$ are the only source of $U(3)^3$ breaking.

\end{enumerate}
All the composite-elementary mixings are then fixed by the known quark masses and CKM angles, up to the parameters $x_t$ (and, in the bidoublet model, $z$), which are defined as in (\ref{eq:x-doublet},\,\ref{eq:x-bidoublet}).
Compared to the anarchic case, one now expects the presence of resonances related to the global symmetry $U(3)_{Q+U+D}$ or $U(3)_{Q^u+U}\times U(3)_{Q^d+D}$, which in the following will be called flavour gauge bosons\footnote{We will only allow flavour gauge bosons related to the $SU(3)$ subgroups of the $U(3)$ factors.} and assumed to have the same masses $m_\rho$ and $g_\rho$ as the gauge resonances. Note that left-compositeness can be meaningfully defined for any of the three cases for the fermion representations, whereas right-compositeness allows to describe flavour violations only in the bidoublet model.

\subsubsection*{$U(2)^3$}

In $U(2)^3$ models one considers a $U(2)_q\times U(2)_u\times U(2)_d$ symmetry, under which the first two generations of quarks transform as doublets and the third generation as singlets, broken in specific directions dictated by minimality \cite{Barbieri:2011ci,Barbieri:2012uh}. Compared to \Utre, one has a larger number of free parameters, but can break the flavour symmetry {\em weakly}, since the large top Yukawa is invariant under \Udue.

Analogously to the \Utre\ case, in the strong sector the flavour groups are $U(2)_{Q+U+D}$ (or $U(2)_{Q^u+U}\times U(2)_{Q^d+D}$) and:

\begin{enumerate}
\item In left-compositeness, to be called {\boldmath\UdueLC}, the left mixings are diagonal with the first two entries equal to each other and the only sources of $U(2)^3$ breaking reside in the right-handed mixings.
\item In right-compositenss, to be called {\boldmath\UdueRC}, the right mixings are diagonal with the first two entries equal to each other and the only sources of $U(2)^3$ breaking reside in the left-handed mixings.
\end{enumerate}
Again we expect the presence of flavour gauge bosons associated with the global symmetries of the strong sector. As before right-compositeness can be meaningfully defined only in the bidoublet model.

\section{General electroweak precision constraints}\label{sec:ewpt}

In this section we discuss electroweak precision constraints that hold independently of the flavour structure. Among the models considered, only \UtreLC\ is subject to additional electroweak constraints, to be discussed in section~\ref{sec:U3EW}.

\subsection{Oblique corrections}

As well known, the $S$ parameter receives a tree-level contribution, which for degenerate composite vectors reads \cite{Contino:2006nn}
\begin{equation}
S = \frac{8 \pi v^2}{m_\rho^2} ~,
\label{S-corr}
\end{equation}
independently of the choice of fermion representations. It is also well known that $S$ and $T$ both get at one loop model-independent ``infrared-log'' contributions \cite{Barbieri:2007bh}
\begin{equation}
\hat{S} =  \left(\frac{v}{f}\right)^2 \frac{g^2}{96\pi^2} \log{\frac{m_\rho}{m_h}}~,
\qquad
 \hat{T} = -   \left(\frac{v}{f}\right)^2 \frac{3 g^2 t^2_w}{32\pi^2} \log{\frac{m_\rho}{m_h}}~.
 \label{inf-logs}
\end{equation}
where $\hat{S} = \alpha_\text{em}/(4 s^2_w)S$ and $\hat{T}=\alpha_\text{em}T$.

Experimentally, a recent global electroweak fit after the discovery of the Higgs boson \cite{Baak:2012kk} finds $S - S_{\text{SM}}=0.03\pm0.10$ and $T- T_{\text{SM}}=0.05\pm0.12$. Requiring $2\sigma$ consistency with these results of the tree level correction to $S$, eq.~(\ref{S-corr}), which largely exceeds the infrared logarithmic contribution of (\ref{inf-logs}) and has the same sign, gives the bound
\begin{equation}
m_\rho>2.6\,\text{TeV} \,.
\label{eq:boundmrho}
\end{equation}

The one loop correction to the $T$ parameter instead strongly depends on the choice of the fermion representations. 
We present here simplified formulae valid in the three models for a common fermion resonance mass $m_\psi$ and developed to first nonvanishing order in $\lambda_{Lt}, \lambda_{Rt}$, as such only valid for small $s_{Lt}, s_{Rt}$. 

In the {doublet model} the leading contribution to $\hat{T}$, proportional to $\lambda_{Rt}^4$, reads
\begin{equation}
\hat{T} = \frac{71}{140} \frac{N_c}{16 \pi^2} \frac{m_t^2}{m_\psi^2} \frac{Y^2}{x_t^2}\,.
\label{eq:T-doublet}
\end{equation}

In the {bidoublet model} one obtains from a leading $\lambda_{Lt}^4$ term 
\begin{equation}
\hat{T} = - \frac{107}{420}\frac{N_c}{16 \pi^2} \frac{m_t^2}{m_\psi^2} x_t^2 Y_U^2\,.
\label{eq:T-bidoublet}
\end{equation}

In the {triplet model} the leading contributions are
\begin{equation}
\hat{T} = \Big(\log\frac{\Lambda^2}{m_\psi^2} - \frac{1}{2}\Big)\frac{N_c}{16 \pi^2} \frac{m_t^2}{m_\psi^2} \frac{Y^3}{y_t x_t}\,, \quad \text{and} \quad \hat{T} = \frac{197}{84}\frac{N_c}{16 \pi^2} \frac{m_t^2}{m_\psi^2} x_t^2 Y^2\,,
\label{eq:T-triplet}
\end{equation}
where the first comes from $\lambda_{Rt}^2$ and the second from $\lambda_{Lt}^4.$ Note the logarithmically divergent contribution to the $\lambda_{Rt}^2$ term that is related to the explicit breaking of the $SU(2)_R$ symmetry in the elementary-composite fermion mixing and would have to be cured in a more complete model.

Imposing the experimental bound at $2\sigma$, eqs.~(\ref{eq:T-doublet}, \ref{eq:T-bidoublet}, \ref{eq:T-triplet}) give rise to the bounds on the first line in table~\ref{tab:bounds-ew} (where we set $\log{(\Lambda/m_\psi)} = 1$). Here however there are two caveats. First, as mentioned, eqs.~(\ref{eq:T-doublet}, \ref{eq:T-bidoublet}, \ref{eq:T-triplet})  are only valid for small mixing angles. Furthermore, for moderate values of $f$, a cancellation could take place between the fermionic contributions and the infrared logs of the bosonic contribution to $T$.
As we shall see, the bounds  from $S$ and $T$ are anyhow not the strongest ones that we will encounter: they are compatible with  $m_\psi\lesssim 1$ TeV for $Y = 1$ to $2$ and $g_\rho = 3$ to $5$. Note that here and in the following $m_{\psi}$ represents the mass of the composite fermions that mix with the elementary ones, whereas, as already noticed, the ``custodians'' have mass $m_{\psi}/\sqrt{1+(\lambda_X)^2}$.

\begin{table}[tbp]
\renewcommand{\arraystretch}{1.3}
\centering
\begin{tabular}{cccc}
\hline
Observable & \multicolumn{3}{c}{Bounds on $m_{\psi}$ [TeV]} \\
& doublet  & triplet & bidoublet \\
\hline
$T$ & $0.28 ~Y/x_t$ &$0.51 ~\sqrt{Y^3/x_t}$, \; $0.60  ~x_t Y$  & $0.25 ~x_tY_U$\\
$R_b$ ($g_{Zbb}^L$) & $5.6  ~\sqrt{x_tY}$ & & $6.5  ~Y_D \sqrt{x_t/Y_{U}}/z$ \\
$B\to X_s\gamma$ ($g_{Wtb}^R$) & $0.44  ~\sqrt{Y/x_t}$  & $0.44  ~\sqrt{Y/x_t}$ & 0.61\\
\hline
\end{tabular}
\caption{Lower bounds on the fermion resonance mass $m_\psi=Y f$ in TeV from electroweak precision observables. A blank space means no significant bound.}
\label{tab:bounds-ew}
\end{table}

\subsection{Modified $Z$ couplings}\label{sec:Zbb}

In all three models for the electroweak structure, fields with different $SU(2)_L$ quantum numbers mix after electroweak symmetry breaking, leading to modifications in $Z$ couplings which have been precisely measured at LEP. Independently of the flavour structure, an important constraint comes from the $Z$ partial width into $b$ quarks, which deviates by $2.5\sigma$ from its best-fit SM value \cite{Baak:2012kk}
\begin{align}
R_b^\text{exp} &= 0.21629(66)~,
&
R_b^\text{SM} &= 0.21474(3)~.
\label{eq:Rbexp}
\end{align}
Writing the left- and right-handed $Z$ couplings as
\begin{equation}
\frac{g}{c_w}\bar b \gamma^\mu \left[
(-\tfrac{1}{2}+\tfrac{1}{3}s_w^2+\delta g^L_{Zbb}) P_L
+(\tfrac{1}{3}s^2_w+\delta g^R_{Zbb})P_R
\right]bZ_\mu
\,,
\end{equation}
one gets
\begin{align}
\delta g_{Zbb}^L &=
\frac{v^2Y_{D}^2}{2m_{D}^2}\frac{xy_t}{Y_U} \,a+
\frac{g_\rho^2v^2}{4m_\rho^2}\frac{xy_t}{Y_U}\,b \,,
&
\delta g_{Zbb}^R &=
\frac{v^2Y_{D}^2}{2m_{D}^2}\frac{y_b^2 Y_U}{x_t y_t Y_D^2} \,c+
\frac{g_\rho^2v^2}{4m_\rho^2}\frac{y_b^2 Y_U}{x_t y_t Y_D^2}\,d \,,
\label{eq:Zbb1}
\end{align}
with the coefficients
\begin{center}
\begin{tabular}{c|ccc}
& doublet & triplet & bidoublet \\
\hline
$a$ & $1/2$   & $0$ & $1/(2 z^2)$\\
$b$ & $1/2$ & $0$ & $1/z^2$
\end{tabular}
\qquad
\begin{tabular}{c|ccc}
& doublet & triplet & bidoublet \\
\hline
$c$ & $-1/2$   & $-1/2$ & $0$\\
$d$ & $-1/2$ & $-1$ & $0$
\end{tabular}
\end{center}
The vanishing of some entries in (\ref{eq:Zbb1}) can be simply understood by the symmetry considerations of ref.~\cite{Agashe:2006at}. 
As manifest from their explicit expressions the contributions proportional to $a$ and $c$ come from mixings between elementary and composite fermions with different $SU(2)\times U(1)$ properties, whereas the contributions proportional to $b$ and $d$ come from $\rho$-$Z$ mixing. Taking $Y_U=Y_D=Y$, $m_D = Yf$ and $m_\rho = g_\rho f$, all these contributions scale however in the same way as $1/(f^2 Y)$.

It is important to note that $\delta g^L_{Zbb}$ is always positive or 0, while $\delta g^R_{Zbb}$ is always negative or 0, while the sign of the SM couplings is opposite. As a consequence, in all 3 models considered, the tension in eq.~(\ref{eq:Rbexp}) is {\em always increased}. Allowing the discrepancy to be at most $3\sigma$, we obtain the numerical bounds in the second row of table~\ref{tab:bounds-ew}. The bound on $m_\psi$ in the doublet model is highly significant since $x_t Y > 1$, whereas it is irrelevant in the triplet model and can be kept under control in the bidoublet model for large enough $z$ (but see below). In the triplet model, there is a bound from the modification of the right-handed coupling, which is however insignificant.

\subsection{Right-handed $W$ couplings}\label{sec:Wtb}

Analogously to the modified $Z$ couplings, also the $W$ couplings are modified after EWSB. Most importantly, a right-handed coupling of the $W$ to quarks is generated. The most relevant experimental constraint on such coupling is the branching ratio of $B\to X_s\gamma$, because a right-handed $Wtb$ coupling lifts the helicity suppression present in this loop-induced decay in the SM \cite{Vignaroli:2012si}. Writing this coupling as
\begin{equation}
\frac{g}{\sqrt{2}}\delta g^R_{Wtb} 
(\bar t \gamma^\mu
P_R
b)W_\mu^+
\,,
\end{equation}
one gets
\begin{align}
\delta g_{Wtb}^R &=
\frac{v^2Y_UY_D}{2m_{Q}m_{U}}\frac{y_b}{x_t Y_U} \,a+
\frac{g_\rho^2v^2}{4m_\rho^2}\frac{y_b}{x_t Y_U}\,b \,,
\label{eq:Wtb1}
\end{align}
with the coefficients
\begin{center}
\begin{tabular}{c|ccc}
& doublet & triplet & bidoublet \\
\hline
$a$ & $1$ & $1$ & $-2x_t y_t/Y$\\
$b$ & $1$ & $1$ & $0$
\end{tabular}
\end{center}
The coefficients in the bidoublet model vanish at quadratic order in the elementary-composite mixings as a consequence of a discrete symmetry \cite{Agashe:2006at}. The nonzero value for $a$ in the table is due to the violation of that symmetry at quartic order \cite{Vignaroli:2012si}.
The contribution to the Wilson coefficient $C_{7,8}$, defined as in \cite{Altmannshofer:2012az}, reads
\begin{equation}
C_{7,8} = \frac{m_t}{m_b} \frac{\delta g^R_{Wtb}}{V_{tb}} A_{7,8}(m_t^2/m_W^2)
\end{equation}
where $A_7(m_t^2/m_W^2)\approx -0.80$ and $A_8(m_t^2/m_W^2)\approx -0.36$.

Since the $B\to X_s\gamma$ decay receives also UV contributions involving composite dynamics, we impose the conservative bound that the SM plus the IR contributions above do not exceed the experimental branching ratio by more than $3\sigma$. In this way we find the bound in the last row of table~\ref{tab:bounds-ew}.

\section{Constraints on the anarchic model}\label{sec:anarchy}

We now discuss constraints that are specific to the anarchic model, as defined above, and hold in addition to the bounds described in the previous section.

\subsection{Tree-level $\Delta F=2$ FCNCs}

In the anarchic model exchanges of gauge resonances give rise to $\Delta F=2$ operators at tree level. Up to corrections of order $v^2/f^2$, the Wilson coefficients of the operators
\begin{align}
Q_V^{dLL} &= (\bar d^i_L\gamma^\mu d^j_L)(\bar d^i_L\gamma^\mu d^j_L) \,,
&
Q_V^{dRR} &= (\bar d^i_R\gamma^\mu d^j_R)(\bar d^i_R\gamma^\mu d^j_R) \,,
\\
Q_V^{dLR} &= (\bar d^i_L\gamma^\mu d^j_L)(\bar d^i_R\gamma^\mu d^j_R) \,,
&
Q_S^{dLR} &= (\bar d^i_R d^j_L)(\bar d^i_L d^j_R) \,,
\end{align}
can be written as
\begin{align}
C_D^{dAB} &= \frac{g_\rho^2}{m_\rho^2}
g_{Ad}^{ij}g_{Bd}^{ij}
c_D^{dAB},& A,B &= L,R,\quad D = V,S,
\label{eq:DF2}
\end{align}
and with the obvious replacements for up-type quarks, relevant for $D$-$\bar D$ mixing.

The couplings $g_{qA}^{ij}$ with $i\neq j$ contain two powers of elementary-composite mixings. In the doublet and triplet models, one can use eqs.~(\ref{eq:anarchy-yukawas})--(\ref{eq:x-doublet}) to write them as ($\xi_{ij} = V_{tj}V_{ti}^*$)
\begin{align}
g_L^{ij}&\sim s_{Ldi}s_{Ldj}\sim \xi_{ij} \frac{x_t y_t}{Y} \,,
\\
g_{Ru}^{ij}& \sim s_{Rui}s_{Ruj}\sim \frac{y_{u^i}y_{u^j}}{Y y_t x_t \xi_{ij}} \,,
\\
g_{Rd}^{ij}& \sim s_{Rdi}s_{Rdj}\sim \frac{y_{d^i}y_{d^j}}{Y y_t x_t \xi_{ij}} \,.
\end{align}
In the bidoublet model,
one has
\begin{align}
g_{Ld}^{ij}\sim g_{Lu}^{ij}&\sim \xi_{ij} \frac{x_t y_t}{Y_U} \,,
&
g_{Rd}^{ij}&\sim z^2\frac{Y_U}{Y_D^2}\frac{y_{d^i}y_{d^j}}{y_t x_t \xi_{ij}} \,.
&
g_{Ru}^{ij}&\sim \frac{y_{u^i}y_{u^j}}{Y_Uy_t x_t \xi_{ij}} \,.
\end{align}
The coefficients $c_D^{AB}$ are discussed in appendix~\ref{sec:app-flavour}.

The experimental bounds on the real and imaginary parts of the Wilson coefficients have been given in \cite{Isidori:2011qw,Calibbi:2012at}.
Since the phases of the coefficients can be of order one and are uncorrelated, we derive the bounds assuming the phase to be maximal. We obtain the bounds in the first eight rows of table~\ref{tab:bounds-anarchy}.
As is well known, by far the strongest bound, shown in the first row, comes from the scalar left-right operator in the kaon system which is enhanced by RG evolution and a chiral factor. Note in particular the growth with $z$ of the bound in the bidoublet case, which counteracts the $1/z$ behaviour of the bound from $R_b$. But also the left-left vector operators in the kaon, $B_d$ and $B_s$ systems lead to bounds which are relevant in some regions of parameter space. The bounds from the $D$ system are subleading.

\begin{table}[tbp]
\renewcommand{\arraystretch}{1.3}
\centering
\begin{tabular}{clll}
\hline
Observable & \multicolumn{3}{c}{Bounds on $m_{\psi}$ [TeV]} \\
& doublet & triplet & bidoublet \\
\hline
$\epsilon_K$ $(Q_S^{LR})$ & $14 $  & $14 $& $14  ~z $ \\
$\epsilon_K$ $(Q_V^{LL})$ & $2.7 ~x_t$& $3.9  ~x_t$  & $3.9  ~x_t$ \\
$B_d$-$\bar B_d$ $(Q_S^{LR})$ & $0.7 $ & $0.7  $ & $0.7 $ \\
$B_d$-$\bar B_d$ $(Q_V^{LL})$ & $2.3 ~x_t$ & $3.4  ~x_t$ & $3.4  ~x_t$ \\
$B_s$-$\bar B_s$ $(Q_S^{LR})$ & $0.6 $ & $0.6  $ & $0.6 $ \\
$B_s$-$\bar B_s$ $(Q_V^{LL})$ & $2.3 ~x_t$ & $3.4  ~x_t$ & $3.4  ~x_t$ \\
$D$-$\bar D$ $(Q_S^{LR})$ & $0.5 $ & $0.5  $ & $0.5 $ \\
$D$-$\bar D$ $(Q_V^{LL})$ & $0.4 ~x_t$ & $0.6  ~x_t$ & $0.6  ~x_t$ \\
$K_L\to\mu\mu$ ($f$--$\psi$) & $0.56  ~\sqrt{Y/x_t}$  & $0.56  ~\sqrt{Y/x_t}$ &\\
$K_L\to\mu\mu$ ($Z$--$\rho$) & $0.39  ~\sqrt{Y/x_t}$  & $0.56  ~\sqrt{Y/x_t}$ &\\
\hline
\end{tabular}
\caption{Flavour bounds on the fermion resonance mass $m_\psi$ in TeV in the anarchic model.}
\label{tab:bounds-anarchy}
\end{table}

\subsection{Flavour-changing $Z$ couplings}

Similarly to the modified flavour-conserving $Z$ couplings discussed in section~\ref{sec:Zbb}, also {\em  flavour-changing} $Z$ couplings are generated in the anarchic model. In the triplet and doublet models, as well as in the bidoublet model, since the down-type contributions to the CKM matrix are not smaller than the up-type contributions in \eqref{eq:anarchy-ckm-bidoublet}, one has
\begin{gather}
\delta g_{Zd^id^j}^L \sim \frac{s_{Ldi}s_{Ldj}}{s_{Lb}^2} ~\delta g_{Zbb}^L \sim \xi_{ij} ~ \delta g_{Zbb}^L~, 
\label{eq:ZbsL}
\\
\delta g_{Zd^id^j}^R \sim \frac{s_{Rdi}s_{Rdj}}{s_{Rb}^2}~ \delta g_{Zbb}^R \sim \frac{y_{d^i}y_{d^j}}{y_b^2 \xi_{ij}} ~ \delta g_{Zbb}^R~.
\label{eq:ZbsR}
\end{gather}

In the $b\to s$ case, a global analysis of inclusive and exclusive $b\to s\ell^+\ell^-$ decays \cite{Altmannshofer:2012az} finds $|\delta g_{Zbs}^{L,R}|\lesssim 8\times 10^{-5}$, while in the $s\to d$ case, one finds $|\delta g_{Zsd}^{L,R}|\lesssim 6\times 10^{-7}$ from the $K_L\to\mu^+\mu^-$ decay \cite{Buras:2011ph}\footnote{The decay $K^+\to\pi^+\nu\bar\nu$ leads to a bound $|\delta g_{Zsd}^{L,R}|\lesssim 3 \times10^{-6}$ at 95\% C.L. and is thus currently weaker than $K_L\to\mu^+\mu^-$, even though it is theoretically much cleaner.}. Using (\ref{eq:ZbsL}) one finds that the resulting constraints on the left-handed coupling are comparable for $b\to s$ and $s \to d$. Since they are about a factor of 3 weaker than the corresponding bound from $Z\to b\bar b$, we refrain from listing them in table~\ref{tab:bounds-anarchy}, but their presence shows that the strong bound from $R_b$ cannot simply be circumvented by a fortuitous cancellation.
In the case of the right-handed coupling, one finds that the constraint from $K_L\to\mu^+\mu^-$ is an order of magnitude stronger than the one from $b\to s\ell^+\ell^-$, and also much stronger than the bound on the right-handed coupling coming from $Z\to b\bar b$. The numerical bounds we obtain are shown in the last two rows of table~\ref{tab:bounds-anarchy} from the contributions with fermion or gauge boson mixing separately since, in constrast to $Z\to b\bar b$, the two terms are multiplied by different $O(1)$ parameters in the flavour-violating case.

\subsection{Loop-induced chirality-breaking effects}

Every flavour changing effect discussed so far originates from tree-level chirality-conserving interactions of the vector bosons, either the elementary $W_\mu$ and $Z_\mu$ or the composite $\rho_\mu$. At loop level, chirality-breaking interactions occur as well, most notably with the photon and the gluon, which give rise  in general to significant  $\Delta F=1$ flavour-changing effects ($b\rightarrow s \gamma$, $\epsilon_K^\prime$, $\Delta A_{CP}(D\rightarrow PP)$), as well as to electric dipole moments of the light quarks. In the weak mixing limit between the elementary and the composite fermions, explicit calculations of some of the  $\Delta F=1$ effects have been made in \cite{Agashe:2008uz,Vignaroli:2012si,Gedalia:2009ws}, obtaining bounds in the range $m_\psi > (0.5\text{--}1.5)Y$\,TeV. For large CP-violating phases the generated EDMs for the light quarks can be estimated  consistent with the current limit on the neutron EDM only if $m_\psi > (3\text{--}5)Y$\,TeV, where the limit is obtained from the 
analysis of \cite{Barbieri:2012bh}. 

\subsection{Direct bounds on vector resonances}\label{sec:anarchy-jjres}

Direct production of vector resonances and subsequent decay to light quarks can lead to a peak in the invariant mass distribution of $pp\to jj$ events at the LHC. In the anarchic model, due to the small degree of compositeness of first generation quarks, the coupling of vector resonances to a first generation quark-antiquark pair is dominated by mixing with the SM gauge bosons and thus suppressed by $g_\text{el}^2/g_\rho$. For a 3~TeV gluon resonance at the LHC with $\sqrt{s}=8$~TeV, following the discussion in appendix~\ref{sec:app-dijet} we expect
\begin{equation}
\sigma(pp\to G^*) = \frac{2\pi}{9s}\frac{g_3^4}{g_\rho^2}
\left[\mathcal L_{u\bar u}(s,m_\rho^2) +\mathcal L_{d\bar d}(s,m_\rho^2)\right]
\approx
\frac{5 ~\text{fb}}{g_\rho^2}\,.
\end{equation}
The ATLAS collaboration has set an upper bound of 7~fb on the cross section times branching ratio to two jets times the acceptance \cite{ATLAS-CONF-2012-088}, and a similar bound has been obtained by CMS \cite{CMS-PAS-EXO-12-016}. Given that the gluon resonance will decay dominantly to top quarks, we conclude that the bound is currently not relevant, even for small $g_\rho$.

A similar argument holds in the case of the dijet angular distribution, which can be used to constrain local four-quark operators mediated by vector resonances. Following the discussion in appendix~\ref{sec:app-dijet-angular}, we obtain the bound
\begin{equation}
m_\rho > \frac{4.5~\text{TeV}}{g_\rho}
\end{equation}
which, in combination with the bound on $m_\rho$ from the $S$ parameter, is irrelevant for $g_\rho\gtrsim1.5$.

\subsection{Partial summary and prospects on anarchy}

If the bound coming from the $Q_S^{LR}$ contribution to $\epsilon_K$ is taken at face value, the fermion resonances should be far too heavy to be consistent with a naturally light Higgs boson and certainly unobservable, either directly or indirectly. Note in particular the growth of this bound with $z$ in the bidoublet model.

In view of the fact that this bound carries an $O(1)$ uncertainty, one might however speculate on what happens if this constraint is ignored.
As visible from table~\ref{tab:bounds-anarchy}, with the exception of the first line, all the strongest bounds  on $m_\psi$ in the bidoublet or in the triplet models can be reduced down to about 1 TeV by taking $x_t = \frac{1}{3}$ to $\frac{1}{4}$. This however correspondingly requires $Y = 3$ to $4$ (and maximal right-handed mixing) which pushes up the bounds from $K_L\rightarrow \mu^+ \mu^-$ and is not consistent with $m_\psi = Yf$ and $f \gtrsim 0.5$~TeV. The loop-induced chirality-breaking effects on $\epsilon^\prime$ and $\Delta A_{CP}$ in $D\rightarrow P P$ decays would also come into play. Altogether, even neglecting the bound from $\epsilon_K(Q_S^{LR})$, fermion resonances below about 1.5 TeV seem hard to conceive.

\section{Constraints on $U(3)^3$}\label{sec:u3}

We now discuss the constraints specific to \Utre. In \UtreLC\ the sizable degree of compositeness of light left-handed quarks leads to additional contributions to electroweak precision observables; in \UtreRC\ FCNCs arise at the tree level. In both cases collider bounds on the compositeness of light quarks place important constraints. Our analysis follows and extends the analysis in \cite{Redi:2011zi}.

\subsection{Electroweak precision constraints specific to $U(3)^3$}\label{sec:U3EW}

The bounds from $R_b$ as well as the $S$ and $T$ parameters discussed in section~\ref{sec:ewpt} are also valid in $U(3)^3$, with one modification: in  \UtreLC\ the contributions to the $\hat T$ parameter proportional to $s_{Lt}^4$ have to be multiplied by 3 since all three generations of left-handed up-type quarks contribute. The corresponding
bounds remain nevertheless relatively mild.

In addition, an important constraint arises from the partial width of the $Z$ into hadrons normalized to the partial width into leptons, which was measured precisely at LEP
\begin{align}
R_h^\text{exp} &= 20.767(25)~,
&
R_h^\text{SM} &= 20.740(17)~,
\end{align}
showing a $1.1\sigma$ tension with the best-fit SM prediction \cite{Baak:2012kk}.

In \UtreLC\ the modified left-handed $Z$ couplings of up and down quarks are equal to the ones of the $t$ and $b$ respectively, while the same is true in \UtreRC\ for the right-handed modified couplings. Analogously to the discussion in section \ref{sec:Zbb}, one can write the modified $Z$ coupling of the top as
\begin{equation}
\frac{g}{c_w}\bar t \gamma^\mu \left[
(\tfrac{1}{2}-\tfrac{2}{3}s_w^2+\delta g^L_{Ztt}) P_L
+(-\tfrac{2}{3}s^2_w+\delta g^R_{Ztt})P_R
\right]tZ_\mu
\,,
\end{equation}
and one has
\begin{align}
\delta g_{Ztt}^L &=
\frac{v^2Y_{U}^2}{2m_{U}^2}\frac{x_ty_t}{Y_U} \,a+
\frac{g_\rho^2v^2}{4m_\rho^2}\frac{x_ty_t}{Y_U}\,b \,,
&
\delta g_{Ztt}^R &=
\frac{v^2Y_{U}^2}{2m_{U}^2}\frac{y_t}{x_t Y_U} \,c+
\frac{g_\rho^2v^2}{4m_\rho^2}\frac{y_t}{x_t Y_U}\,d \,,
\label{eq:Zbb}
\end{align}
with
\begin{center}
\begin{tabular}{c|ccc}
& doublet & triplet & bidoublet \\
\hline
$a$ & $-1/2$   & $-1$ & $-1/2$\\
$b$ & $-1/2$ & $-1$ & $-1$
\end{tabular}
\qquad
\begin{tabular}{c|ccc}
& doublet & triplet & bidoublet \\
\hline
$c$ & $1/2$   & $0$ & $0$\\
$d$ & $1/2$ & $0$ & $0$
\end{tabular}
\end{center}
Since the right-handed $Z$ coupling to $b$ and $t$ receives no contribution in the bidoublet model, there is no additional bound from $R_h$ in \UtreRC.
In \UtreLC we find the numerical bounds shown in table~\ref{tab:bounds-U3LC}.

In \UtreLC\ an additional bound arises from violations of quark-lepton universality.
Writing the $W$ couplings as
\begin{equation}
\frac{g}{\sqrt{2}}(1+\delta g^L_{W}) \bar u\, V_{ui}\gamma^\mu 
P_L
d_i W_\mu^+
\,,
\end{equation}
we find
\begin{align}
\delta g_{W}^L &=
\frac{Y_{U}^2v^2}{2m_{U}^2}\frac{x_ty_t}{Y_U} \,a_u +
\frac{Y_{D}^2v^2}{2m_{D}^2}\frac{x_ty_t}{Y_U} \,a_d +
\frac{g_\rho^2v^2}{4m_\rho^2}\frac{x_ty_t}{Y_U}\,b \,,
\end{align}
with
\begin{center}
\begin{tabular}{c|ccc}
& doublet & triplet & bidoublet \\
\hline
$a_u$ & $-1/2$ & $-1/2$ & $-1/2$
\\
$a_d$ & $-1/2$ & $-1/2$ & $-1/(2z^2)$
\\
$b$ & $-1$ & $-1$ & $-1$
\end{tabular}
\end{center}
The usual experimental constraint on the strength of the $W\bar{u}d_i$ couplings, normalized to the leptonic ones, is expressed by $(1+\delta g^L_{W})^2 \sum_i|V_{ui}|^2-1=(-1\pm6)\times10^{-4}$, which, from the unitarity of the $V_{ij}$ matrix, becomes $2 \delta g^L_{W}= (-1\pm6)\times10^{-4}$.  By requiring it to be fulfilled within $2\sigma$, we find the numerical bounds in table~\ref{tab:bounds-U3LC}.

Finally we note that, in contrast to the anarchic case, there are no {\em flavour-changing} $Z$ couplings neither in \UtreLC\ nor in \UtreRC. In the former case this is a general property of chirality-conserving bilinears, while in the latter it is a consequence of the fact that only the down-type mixings $\lambda_{Ld}$ affect the $Z$ vertex, which thus becomes flavour-diagonal in the mass basis.

\subsection{Tree-level  $\Delta F=2$ FCNCs}

While in \UtreLC\ there are no tree-level FCNCs at all \cite{Redi:2011zi}, minimally flavour violating tree-level FCNCs are generated in \UtreRC\ \cite{Barbieri:2012uh,Redi:2012uj} (see appendix~\ref{sec:app-U2U3}). The Wilson coefficients of $\Delta F=2$ operators are given by (\ref{eq:DF2}), with the coefficients $c_D^{qAB}$ listed in appendix~\ref{sec:app-flavour} and the couplings 
\begin{align}
g_{Ld}^{ij} &= \xi_{ij}\frac{x_ty_t}{Y_U} \,,
&
g_{Rd}^{ij} &\approx 0 \,.
\label{U(3)RC-FC}
\end{align}
We obtain the bounds shown in table~\ref{tab:bounds-U3RC}. The bound from $D$-$\bar D$ mixing turns out to be numerically irrelevant.

We stress that, in contrast to the anarchic case, there is no $O(1)$ uncertainty in these bounds since the composite Yukawas are proportional to the identity. Furthermore, since the model is minimally flavour violating, there is no contribution to the meson mixing phases and the new physics effects in the $K$, $B_d$ and $B_s$ systems are prefectly correlated.

\begin{table}[tbp]
\renewcommand{\arraystretch}{1.3}
\centering
\begin{tabular}{cccc}
\hline
Observable & \multicolumn{3}{c}{Bounds on $m_{\psi}$ [TeV]} \\
&  doublet & triplet  & bidoublet\\
\hline
$R_h$ & $7.2 ~\sqrt{x_tY}$  &$6.0 ~\sqrt{x_tY}$  &$4.9 ~\sqrt{x_tY_U}$\\
$V_\text{CKM}$ & $7.4 ~\sqrt{x_tY}$ & $7.4 ~\sqrt{x_tY}$ & $6.0 ~\sqrt{x_tY_U}$ \\
$pp\to jj$ ang. dist. & $3.4  ~x_t$ & $4.2  ~x_t$ & $4.2  ~x_t$  \\
\hline
\end{tabular}
\caption{Lower bounds on the fermion resonance mass $m_\psi$ in TeV in \UtreLC.}
\label{tab:bounds-U3LC}
\end{table}

\begin{table}[tbp]
\renewcommand{\arraystretch}{1.3}
\centering
\begin{tabular}{cc}
\hline
Observable & Bounds on $m_{\psi}$ [TeV] \\
\hline
$\epsilon_K(Q^{LL}_V)$ & $3.7 ~x_t$ \\
$B_d$-$\bar B_d$ & $3.2 ~x_t$ \\
$B_s$-$\bar B_s$  & $3.6 ~x_t$\\
$pp\to jj$ ang. dist. & $3.0/x_t$ \\
\hline
\end{tabular}
\caption{Lower bounds on the fermion resonance mass $m_\psi$ in TeV in \UtreRC\ (bidoublet model).}
\label{tab:bounds-U3RC}
\end{table}

\subsection{Loop-induced chirality-breaking effects}

Flavour-changing chirality-breaking effects in $U(3)^3$ occur when elementary-composite mixings are included inside the loops. At least for moderate mixings, the bounds are of the form $m_\psi > (0.5\text{--}1.5) \sqrt{Y/x_t} $ TeV in the \UtreLC\ case, or $m_\psi > (0.5\text{--}1.5)\sqrt{Y x_t}$ TeV in the \UtreRC\ case. The stronger bounds from  quark EDMs, similar to the ones of the anarchic case,  disappear if the strong sector conserves CP. This is automatically realized, in our effective Lagrangian description, if the ``wrong chirality'' Yukawas vanish or are aligned in phase with the $Y$'s. On the contrary, in the anarchic case this condition is in general not sufficient to avoid large EDMs.

\subsection{Compositeness constraints}

Since one chirality of first-generation quarks has a sizable degree of compositeness in the $U(3)^3$ models, a significant constraint arises from the angular distribution of dijet events at LHC, which is modified by local four-quark operators obtained after integrating out the heavy vector resonances related to the global $SU(3)_c\times SU(2)_L\times SU(2)_R\times U(1)_X$ as well as the flavour symmetry in the strong sector, $U(3)$ in the case of \UtreLC\ and $U(3)\times U(3)$ in the case of \UtreRC.

In general, there are ten four-quark operators relevant in the dijet angular distribution \cite{Domenech:2012ai}. Following the discussion in appendix~\ref{sec:app-dijet-angular}, the relevant operators in \UtreLC\ are $\mathcal O_{qq}^{(1,8)}$. Their Wilson coefficients read
\begin{align}
C_{qq}^{(1)} = -\frac{a}{36}\frac{g_\rho^2}{m_\rho^2}\left(\frac{x_t y_t}{Y_U}\right)^2\,,
\qquad
C_{qq}^{(8)} = -\frac{g_\rho^2}{m_\rho^2}\left(\frac{x_t y_t}{Y_U}\right)^2\,,
\end{align}
where $a=5$ in the doublet model and $a=17$ in the triplet and bidoublet models.
Using the updated version of \cite{Domenech:2012ai},
we obtain the bound in the last row of table~\ref{tab:bounds-U3LC}.
In \UtreRC\ the operators with right-handed quarks are relevant, i.e. $\mathcal O_{uu,dd,ud}^{(1)}$ and $\mathcal O_{ud}^{(8)}$.
Numerically, we find the bound on $\mathcal O_{uu}^{(1)}$ to give the most significant constraint on the model parameters. Its Wilson coefficient reads
\begin{align}
C_{uu}^{(1)} = -\frac{5}{9}\frac{g_\rho^2}{m_\rho^2}\left(\frac{y_t}{x_t Y_U}\right)^2\,.
\end{align}
and the resulting numerical constraint is shown in the last row of table~\ref{tab:bounds-U3RC}.

\subsection{Direct bounds on vector resonances}\label{sec:u3-jjres}

As discussed in section~\ref{sec:anarchy-jjres}, direct bounds on $m_\rho$ are obtained from searches for peaks in the invariant mass of dijets at LHC. In $U(3)^3$ the production cross sections can be larger than in the anarchic case due to the sizable degree of compositeness of first-generation quarks. Neglecting the contribution due to mixing of the vector resonances with the gauge bosons, the production cross section of a gluon resonance reads (see appendix~\ref{sec:app-dijet})
\begin{align}
\sigma(pp\to G^*) &=  \frac{2\pi}{9s}g_\rho^2
\left[
s_{L,Ru}^4 \mathcal L_{u\bar u}(s,m_\rho^2)
+
s_{L,Rd}^4 \mathcal L_{d\bar d}(s,m_\rho^2)
\right],
\label{eq:sigmapprhoU3}
\end{align}
where the $L$ is valid in \UtreLC\ and the $R$ in \UtreRC.
In \UtreLC\ the branching ratio to two jets reads approximately
\begin{align}
\text{BR}(G^*\to jj) =
\frac{2 s_{Lu}^4+3 s_{Ld}^4+s_{Rb}^4}{3 s_{Lu}^4+s_{Rt}^4+3 s_{Ld}^4+s_{Rb}^4
}\,,
\end{align}
and is typically larger than in the anarchic case. Similarly, in \UtreRC\ one has
\begin{equation}
\text{BR}(G^*\to jj) =
\frac{
2 s_{Ru}^4
+
s_{Lb}^4
+
3 s_{Rd}^4
}{
s_{Lt}^4
+
3 s_{Ru}^4
+
s_{Lb}^4
+
3 s_{Rd}^4
}\,.
\end{equation}

To judge if the most recent experimental bounds by ATLAS \cite{ATLAS-CONF-2012-088} and CMS \cite{CMS-PAS-EXO-12-016} have already started to probe the $U(3)^3$ parameter space, we evaluate the cross section for maximal mixing, i.e. $x_t=Y/y_t$ in \UtreLC\ and $x_t=y_t/Y$ in \UtreRC, for a 3~TeV gluon resonance, i.e. only marginally heavier than allowed by the $S$ parameter (cf. table~\ref{tab:bounds-ew}). For \UtreLC\ we obtain 
\begin{align}
\sigma(pp\to G^*) \approx 13g_\rho^2 ~\text{fb}~,
\qquad
\text{BR}(G^*\to jj) \approx 58\% ~(83\%) \text{ for } Y=1 ~(4\pi)~,
\end{align}
and for \UtreRC
\begin{align}
\sigma(pp\to G^*) \approx 30g_\rho^2 ~\text{fb}~,
\qquad
\text{BR}(G^*\to jj) \approx 69\% ~(67\%) \text{ for } Y=1 ~(4\pi)~.
\end{align}
This is to be compared to the ATLAS bound $\sigma\times\text{BR}\times A<7~\text{fb}$, where $A$ is the acceptance. We conclude that, assuming an acceptance of the order of 60\% \cite{ATLAS-CONF-2012-088}, maximal mixing is on the border of exclusion in \UtreLC\ and already excluded in \UtreRC\ for a 3~TeV gluon resonance.
We note however that maximal mixing is already disfavoured by the indirect bounds discussed above.

\subsection{Partial summary on $U(3)^3$}

As apparent from tables~\ref{tab:bounds-U3LC} and \ref{tab:bounds-U3RC}, a fermion resonance at about 1 TeV is disfavoured.
In \UtreLC\ the crucial constrains come from the EWPT due to the large mixing of the first generations quarks in their left component. Note that $x_t Y$ cannot go below $y_t\sim1$. In  \UtreRC\ there is a clash between the tree-level $\Delta F=2$ FCNC effects, which decrease with $x_t$, and the bound from the $pp\to jj$  angular distributions due to the composite nature of the light quarks in their right component, which goes like $1/x_t$.
We stress again that these conclusions are more robust than in the anarchic case, since there is no uncertainty related to the composite Yukawas, which are flavour universal in the \Utre\ case.

\section{Constraints on $U(2)^3$}\label{sec:u2}

In \UdueLC\ and \UdueRC\ the first and second generation elementary-composite mixings are expected to be significantly smaller than the third generation ones, so that the electroweak 
precision constraints and the collider phenomenology are virtually identical to the anarchic case and the most serious problems plaguing the \Utre\ models are absent. The most important difference concerns the flavour constraints.

\subsection{Tree-level  $\Delta F=2$ FCNCs}

As shown in appendix \ref{sec:app-U2}, the Wilson coefficients of $\Delta F=2$ operators generated in \UdueLC\ and \UdueRC\ are again given by (\ref{eq:DF2}). The flavour-changing couplings in \UdueLC\ read
\begin{align}
g_{Ld}^{i3} &= \xi_{i3} \, r_b\frac{x_t y_t}{Y_U} \,,
&
g_{Ld}^{12} &= \xi_{12} \, |r_b|^2 \frac{x_t y_t}{Y_U} \,,
&
g_{Rd}^{ij} &\approx 0 \,,
\label{FCU2LC}
\end{align}
where $r_b$ is a free complex parameter defined in \eqref{eq:rb}. As a consequence there is a new, universal phase in $B_d$ and $B_s$ mixing, while the $K$-$\bar K$ amplitude is aligned in phase with the SM. We find the bounds in table~\ref{tab:bounds-U2LC}. If the parameter $|r_b|$ is somewhat less than 1, these bounds can be in agreement with experiment even for light fermion resonances. We note that the contribution to the $\Delta C=2$ operator is proportional to $|1-r_b|^2$, so it cannot be reduced simultaneously. However, it turns out that it is numerically insignificant. Since furthermore the contribution is real -- a general prediction of the \Udue\ symmetry for $1\leftrightarrow 2$ transitions -- the expected improvement of the bound on CP violation in $D$-$\bar D$ mixing will have no impact.

In \UdueRC\ the flavour-changing couplings are the same as in \UtreRC,
\begin{align}
g_{Ld}^{i3} &= \xi_{i3} \,\frac{x_ty_t}{Y_U} \,,
&
g_{Ld}^{12} &= \xi_{12} \, \frac{x_t y_t}{Y_U} \,,
&
g_{Rd}^{ij} &\approx 0 \,.
\end{align}
Thus, as in \UtreRC, there is no new phase in meson-antimeson mixing and the NP effects in the $K$, $B_d$ and $B_s$ systems are perfectly correlated. The resulting bounds are shown in table~\ref{tab:bounds-U2RC}.

\subsection{Loop-induced chirality-breaking effects}

One expects in general  flavour-changing chirality-breaking effects in $U(2)^3$ with bounds on the fermion resonances similar to the one of the anarchic case,  $m_\psi > (0.5\text{--}1.5)Y$\,TeV.
With CP conservation in the strong sector, however, the contributions to the quarks EDMs would arise only at higher orders in the $U(2)^3$ breaking terms (see appendix~\ref{sec:app-U2}), so that they would not be significant for the current limit on the neutron EDM.

\begin{table}[tbp]
\renewcommand{\arraystretch}{1.3}
\centering
\begin{tabular}{cccc}
\hline
Observable & \multicolumn{3}{c}{Bounds on $m_{\psi}$ [TeV]} \\
& doublet & triplet& bidoublet  \\
\hline
$\epsilon_K(Q^{LL}_V)$ & $2.3 ~x_t|r_b|^2$ & $3.3  ~x_t|r_b|^2$ & $3.3  ~x_t|r_b|^2$ \\
$B_d$-$\bar B_d$ & $2.3 ~x_t|r_b|$ & $3.4  ~x_t|r_b|$ & $3.4  ~x_t|r_b|$ \\
$B_s$-$\bar B_s$ & $2.3 ~x_t|r_b|$ & $3.4  ~x_t|r_b|$ & $3.4  ~x_t|r_b|$ \\
$K_L\to\mu\mu$ & $3.8 ~\sqrt{x_tY}|r_b|$ && $3.8  ~Y_D |r_b| \sqrt{x_t/Y_{U}}/z$ \\
$b\to s\ell\ell$ & $3.5 ~\sqrt{x_tY|r_b|}$ && $3.5  ~Y_D \sqrt{x_t|r_b|/Y_{U}}/z$ \\
\hline
\end{tabular}
\caption{Lower bounds on the fermion resonance mass $m_\psi$ in TeV in \UdueLC.}
\label{tab:bounds-U2LC}
\end{table}

\begin{table}[tbp]
\renewcommand{\arraystretch}{1.3}
\centering
\begin{tabular}{cc}
\hline
Observable & Bounds on $m_{\psi}$ [TeV] \\
\hline
$\epsilon_K(Q^{LL}_V)$ & $3.3 ~x_t$ \\
$B_d$-$\bar B_d$ & $2.8 ~x_t$ \\
$B_s$-$\bar B_s$  & $3.1 ~x_t$\\
\hline
\end{tabular}
\caption{Lower bounds on the fermion resonance mass $m_\psi$ in TeV in \UdueRC\ (bidoublet model).}
\label{tab:bounds-U2RC}
\end{table}

\subsection{Flavour-changing $Z$ couplings}

In \UdueRC\ flavour-changing $Z$ couplings are absent at tree level. In \UdueLC\ the left-handed couplings do arise, while the right-handed couplings are strongly suppressed. Similarly to the anarchic case, one can write them as
\begin{gather}
\delta g_{Zbd^i}^L \sim  \xi_{i3} ~ r_b ~ \delta g_{Zbb}^L~,
\qquad
\delta g_{Zsd}^L \sim  \xi_{12} ~ |r_b|^2 ~ \delta g_{Zbb}^L~.
\label{eq:ZbsL-U2}
\end{gather}
One obtains the bounds in the last two lines of table~\ref{tab:bounds-U2LC}, which are weaker than the analogous bounds from $R_b$ unless $|r_b|>1$.
An important difference with respect to the anarchic case is the absence of sizable flavour-changing {\em right-handed} $Z$ couplings, which can be probed e.g. in certain angular observables in $B\to K^*\mu^+\mu^-$ decays \cite{Altmannshofer:2008dz}.

\subsection{Partial summary on $U(2)^3$}

Two important differences distinguish the $U(2)^3$ case from the $U(3)^3$ one: i) both for the bidoublet (at  large enough $z$) and for the triplet models, the  bounds from the EWPT or from compositeness become irrelevant; ii) a single complex parameter correlates the various observables, $r_b$ in the  \UdueLC\  case. As apparent from table~{\ref{tab:bounds-U2LC}, values of $x_t$ and $r_b$ somewhat smaller than one can reduce the bounds on the fermion resonance mass at or even below the 1 TeV level. This is also formally possible in \UdueRC, where $r_b=1$, but requires $x_t \lesssim 0.3$, i.e. $Y \gtrsim 3$, not consistent with $m_\psi = Y f$ and $f \gtrsim 0.5$~TeV.

\section{Summary and Conclusions}\label{sec:summary}

After about two years of operation of the LHC and the remarkable discovery of a Higgs-like particle of 125 GeV mass, the view of a natural Fermi scale is still under scrutiny, with three different lines of investigation:  the more precise measurements of the properties of the same Higgs-like boson, the direct searches of new particles that are expected to accompany the Higgs boson and several measurements in flavour physics. One  way to implement a natural Fermi scale is to make the Higgs particle, one or more, a pseudo-Goldstone boson of a new strong interaction in the few TeV range. 
A meaningful question is then if and how a Higgs boson of 125 GeV mass fits into this picture, which requires spin-$\frac{1}{2}$ resonances, partners of the top, with a semi-perturbative coupling to the strong sector and a mass not exceeding about 1 TeV. 

Not the least difficulty in addressing this question is the variety of  possible specific implementations of the Higgs-as-pseudo-Goldstone-boson picture, especially with regard to the different representations of the spin-$\frac{1}{2}$ resonances and the various ways to describe flavour. A further problem is represented by the limited calculability of key observables in potentially complete models, due to their strongly interacting nature. 

\begin{table}[tbp]
\renewcommand{\arraystretch}{1.3}
\centering
\begin{tabular}{cccc}
\hline
& doublet & triplet & bidoublet \\
\hline
\CircledA & $4.9$ & $1.7$ & $1.2*$ \\
\UtreLC & $6.5$ & $6.5$ & $5.3$ \\
\UtreRC &-&-& $3.3$ \\
\UdueLC & $4.9$ & $0.6$ & $0.6$ \\
\UdueRC &-&-& $1.1*$ \\
\hline
\end{tabular}
\caption{Minimal fermion resonance mass $m_\psi$ in TeV  compatible with all the bounds (except for the $Q_S^{LR}$ contribution to $\epsilon_K$ in the anarchic model), fixing $O(1)$ parameters in anarchy to 1 and assuming the parameter $|r_b|$ in \UdueLC\ to be $\sim 0.2$. The bounds with a $*$ are obtained for a  value of  $Y \approx 2.5$, that minimizes the flavour and EWPT constraints consistently with $m_\psi = Y f$ and $f \gtrsim 0.5$ TeV.}
\label{tab:mmin}
\end{table}

To circumvent these difficulties, we have adopted some simple {\it partial-compositeness} Lagrangians and assumed that they catch the basic phenomenological properties of the theories under consideration. This allows us to consider a grid of various possibilities, represented, although at the risk of being too simplistic, in table~\ref{tab:mmin}, which tries to summarize   all in one go the content of the more detailed tables \ref{tab:bounds-ew} to \ref{tab:bounds-U2RC} discussed throughout the paper, taking into account all constraints from flavour and EWPT. For any given case, this table estimates a lowest possible value for the mass of the composite fermions that mix with the elementary ones and which are heavier than the ``custodians'' by a factor of $\sqrt{1 + (\lambda_X)^2}$. In the case of {\it anarchy} we are neglecting the constraint coming from $\epsilon_K$ (first line of table~\ref{tab:bounds-anarchy}, particularly problematic for the bidoublet model, maybe accidentally suppressed) and the various $O(1)$ factors that plague most of the other flavour observables in table~\ref{tab:bounds-anarchy}.
In every case we also neglect the constraint coming from  one-loop chirality-breaking operators, relevant to direct CP violation both in the $K$ and in the $D$ systems, as well as to the quark electric dipole moments. This is a subject that  deserves further detailed study.

\definecolor{green}{cmyk}{0.5,0,1,0.2}
\definecolor{lightgray}{rgb}{0.7,0.7,0.7}
\newcommand{\si}{{\color{green}\footnotesize$\bigstar$}}
\newcommand{\no}{{\color{lightgray}$\circ$}}

\begin{table}
\renewcommand{\arraystretch}{1.3}
\centering
\begin{tabular}{cccccc}
\hline
& ~\CircledA~ & \UtreLC & \UtreRC & \UdueLC & \UdueRC \\
\hline
$\epsilon_K$, $\Delta M_{d,s}$ & \si & \no & \si & \si & \si \\
$\Delta M_{s}/\Delta M_{d}$ & \si & \no & \no & \no & \no \\
$\phi_{d,s}$ & \si & \no & \no & \si & \no \\
$\phi_s-\phi_d$ & \si & \no & \no & \no & \no \\
$C_{10}$ & \si & \no & \no & \si & \no \\
$C_{10}'$ & \si & \no & \no & \no & \no \\
\hline
$pp\to jj$ & \no & \si & \si & \no & \no \\
$pp\to q'q'$ & \si & \no & \no & \si & \si \\
\hline
\end{tabular}
\caption{Observables where NP effects could show up with realistic experimental and/or lattice improvements in the most favourable cases.}
\label{tab:dna}
\end{table}

The general message that emerges from table~\ref{tab:mmin}, taken at face value,  is pretty clear. To accommodate top partners at or below 1 TeV is often not  possible and requires a judicious choice of the underlying model: an approximate $U(2)^3$ flavour symmetry appears favorite, if not necessary. Note that the bounds with a $*$ (bidoublet model with anarchic or  \UdueRC\  flavour structure) are obtained for a value of $Y \approx 2.5$, that minimizes the flavour and EWPT constraints consistently with $m_\psi = Y f$ and $f \gtrsim 0.5$ TeV. There are two simple reasons for the emergence of $U(2)^3$: i) in common with $U(3)^3$, the suppression of flavour changing effects in four-fermion operators with both left- and right-handed currents, present in the anarchic case; ii) contrary to $U(3)^3$ but as in anarchy, the disentanglement of the properties (their degree of compositeness) of the first and third generation of quarks.

The source of the constraint that plays the dominant role in the various cases is diverse. Sometimes more than one observable gives comparable constraints. This is reflected in table~\ref{tab:dna}, which summarizes 
where possible new physics
effects could show up\footnote{The observables are, from top to to bottom: the direct CP violating parameter in $K$-$\bar K$ mixing and the $B_d$ and $B_s$ mass differences (as well as their ratio), the mixing phases $\phi_d,\phi_s$ in the $B_d$ and $B_s$ systems (as well as their difference), the Wilson coefficient of the axial vector semi-leptonic operator relevant for $b\to s\ell^+\ell^-$ transitions $C_{10}$ and its chirality-flipped counterpart $C_{10}'$, the angular distribution of dijet events at LHC as discussed above and the direct production of fermion resonances at LHC.}(for some observables with more experimental
data, for others if  lattice parameters and/or other theoretical inputs are improved). We keep in this table every possible case even though  some of them, according to table~\ref{tab:mmin},   would have to live with a fine tuned Higgs boson squared mass and, as such, appear less motivated.

The attempt to include many different possibilities, though motivated,  is also a limit of the analysis presented in this work. A next step might consist in selecting a few emerging cases to analyze them in more detail, perhaps going beyond the {\it partial-compositeness}  effective description. For this we think that table~\ref{tab:mmin} offers a useful criterion.
It is in any event important and a priori non trivial that some models with a suitable structure emerge that look  capable of accommodating a 125 GeV Higgs boson without too much fine tuning, i.e. with top partners in an interesting mass range for discovery at the LHC.

\section*{Acknowledgments}

D.S. thanks Roberto Contino for useful discussions. This work was supported by the EU ITN ``Unification in the LHC Era'',
contract PITN-GA-2009-237920 (UNILHC) and by MIUR under contract 2008XM9HLM.
The research of D.S. is supported by the Advanced Grant EFT4LHC of the European Research Council (ERC).

\appendix

\section{Tree-level $\Delta F=2$ FCNCs}\label{sec:app-flavour}

In a model with flavour anarchy in the strong sector, the coefficients defined in eq.~(\ref{eq:DF2}) can be written as
\begin{align}
c_V^{dLL} &= -\frac{1}{6} -\frac{1}{2}\left[X(Q)^2+T_{3L}(Q)^2+T_{3R}(Q)^2\right]
\,,\\
c_V^{dRR} &= -\frac{1}{6} -\frac{1}{2}\left[X(D)^2+T_{3L}(D)^2+T_{3R}(D)^2\right]
\,,\\
c_V^{dLR} &= \frac{1}{6}-\left[X(Q)X(D)+T_{3L}(Q)T_{3L}(D)+T_{3R}(Q)T_{3R}(D)\right]
\,,\\
c_S^{dLR} &= 1
\,,
\end{align}
where the first terms come from heavy gluon exchange and the terms in brackets from neutral heavy gauge boson exchange. $Q$ refers to the charge $-1/3$ fermion mixing with $q$ and $D$ to the charge $-1/3$ fermion mixing with $d_R$. In the bidoublet model, we consider only the contribution from $Q_u$ for simplicity, which is enhanced if $z>1$.
The numerical coefficients relevant for the models discussed above are collected in table~\ref{tab:df2coeffs}.

In $U(3)^3$ there is an additional contribution from flavour gauge bosons. However the only relevant $\Delta F = 2$ operator is $Q_V^{dLL}$ in \UtreRC, for which one obtains $c_V^{dLL} = -{29}/{36}$ instead of the value reported in the table.

In $U(2)^3$, since all the flavour effects are generated by mixing with third generation partners, which are not charged under any of the $U(2)$ flavour groups, there is no relevant additional effect coming from flavour gauge bosons, and the coefficients of table \ref{tab:df2coeffs} are valid.

\begin{table}[t]
\renewcommand{\arraystretch}{1.5}
\centering%
\begin{tabular}{cccc}
\hline
 & doublet & triplet & bidoublet\\
\hline
$c_V^{dLL}$ & $-\frac{11}{36}$ & $-\frac{23}{36}$ & $-\frac{23}{36}$\\
$c_V^{dRR}$ & $-\frac{11}{36}$ & $-\frac{8}{9}$ & $-\frac{2}{9}$\\
$c_V^{dLR}$ & $\frac{5}{36}$ & $-\frac{7}{9}$ & $\frac{7}{18}$\\
$c_S^{dLR}$ & 1 & 1 & 1\\
\hline
\end{tabular}
\caption{Coefficients relevant for $\Delta F = 2$ operators in anarchy and $U(2)^3$.}
\label{tab:df2coeffs}
\end{table}

\section{Compositeness constraints from the dijet angular distribution}\label{sec:app-dijet-angular}

Exchanges of gauge resonances and flavour gauge bosons give rise to four-fermion operators involving only the first generation which contribute to the angular distribution of high-mass dijets at LHC. As shown in \cite{Domenech:2012ai}, only ten operators have to be considered, which we list here for convenience
\begin{align}
\mathcal O^{(1)}_{uu}&=(\bar{u}_R\gamma^\mu u_R)(\bar{u}_R\gamma_\mu u_R)~,
&
\mathcal O^{(1)}_{dd}&=(\bar{d}_R\gamma^\mu d_R)(\bar{d}_R\gamma_\mu d_R)~,
\nonumber\\
\mathcal O^{(1)}_{ud}&=(\bar{u}_R\gamma^\mu u_R)(\bar{d}_R\gamma_\mu d_R)~,
&
\mathcal O^{(8)}_{ud}&=(\bar{u}_R\gamma^\mu T^A u_R)(\bar{d}_R\gamma_\mu T^A d_R)~,
\nonumber\\
\mathcal O^{(1)}_{qq}&=(\bar{q}_L\gamma^\mu q_L)(\bar{q}_L\gamma_\mu q_L)~,
&
\mathcal O^{(8)}_{qq}&=(\bar{q}_L\gamma^\mu T^A q_L)(\bar{q}_L\gamma_\mu T^A q_L)~,
\nonumber\\
\mathcal O^{(1)}_{qu} &= (\bar{q}_{L} \gamma^\mu q_{L}) (\bar{u}_{R} \gamma_{\mu}  u_R)~,
&
\mathcal O^{(8)}_{qu} &= (\bar{q}_{L} \gamma^\mu T^A q_{L}) (\bar{u}_{R} \gamma_{\mu}  T^A u_R)~,
\nonumber\\
\mathcal O^{(1)}_{qd} &= (\bar{q}_{L} \gamma^\mu q_{L}) (\bar{d}_{R} \gamma_{\mu}  d_R)~,
&
\mathcal O^{(8)}_{qd} &= (\bar{q}_{L} \gamma^\mu T^A q_{L}) (\bar{d}_{R} \gamma_{\mu}  T^A d_R)~.
\label{eq:jj-operators}
\end{align}
The coupling of a first generation quark mass eigenstate to a heavy vector resonance receives contributions from fermion composite-elementary mixing as well as vector boson composite-elementary mixing. For example, the coupling of the up quark to the gluon resonance reads
\begin{equation}
\bar u \gamma^\mu T^a\left(
g_\rho s_{Lu}^2 P_L +g_\rho s_{Ru}^2 P_R + \frac{g_3^2}{g_\rho}
\right)u G^*_\mu ~.
\label{eq:gG}
\end{equation}
Neglecting electroweak gauge couplings, one can then write the Wilson coefficients of the above operators as
\begin{equation}
C_{ab}^{(1,8)} = \frac{g_\rho^2}{m_\rho^2}\left[
s_{a}^2
s_{b}^2
c_{ab}^{(1,8)}
+
\left(
\frac{g_3^4}{g_\rho^4}
-(s_{a}^2+s_{b}^2)
\frac{g_3^2}{g_\rho^2}
\right)
d_{ab}^{(1,8)}
\right],
\end{equation}
where $(a,b)=(q,u,d)$ and $s_{u,d}^2\equiv s_{Ru,d}^2$, $s_q^2\equiv s_{L}^2$ (in the bidoublet model, for simplicity we will neglect terms with $s_{Ld}^2$ over terms with $s_{Lu}^2$).
The numerical coefficients $c_{ab}^{(1,8)}$ depend on the electroweak structure and on the flavour group and are collected in table~\ref{tab:dijet} together with the $d_{ab}^{(1,8)}$.

\begin{table}[tbp]
\renewcommand{\arraystretch}{1.5}
\centering
\begin{tabular}{lcccccccccc}
\hline
&$c_{uu}^{(1)}$ & $c_{dd}^{(1)}$ & $c_{ud}^{(1)}$ & $c_{ud}^{(8)}$ &
$c_{qq}^{(1)}$ & $c_{qq}^{(8)}$ & $c_{qu}^{(1)}$ & $c_{qu}^{(8)}$ & $c_{qd}^{(1)}$ & $c_{qd}^{(8)}$ \\
\hline
doublet $U(3)^3_\text{LC}$ & $-\frac{17}{36}$ & $-\frac{17}{36}$ & $-\frac{1}{9}$ & $-1$ & $-\frac{5}{36}$ & $-1$ & $-\frac{13}{36}$ & $-1$ & $-\frac{13}{36}$ & $-1$ \\
triplet $U(3)^3_\text{LC}$& $-\frac{5}{9}$ & $-\frac{19}{18}$ & $-\frac{7}{9}$ & $-1$ & $-\frac{17}{36}$ & $-1$ & $-\frac{7}{9}$ & $-1$ & $-\frac{7}{9}$ & $-1$ \\
bidoublet $U(3)^3_\text{LC}$ &  $-\frac{5}{9}$ & $-\frac{7}{18}$ & $-\frac{1}{9}$ & $-1$ & $-\frac{17}{36}$ & $-1$ & $-\frac{7}{9}$ & $-1$ & $-\frac{1}{9}$ & $-1$ \\
bidoublet $U(3)^3_\text{RC}$ & $-\frac{5}{9}$ & $-\frac{7}{18}$ & $\frac{2}{9}$ & $-1$ & $-\frac{17}{36}$ & $-1$ & $-\frac{7}{9}$ & $-1$ & $\frac{2}{9}$ & $-1$ \\
\hline
\hline
&$d_{uu}^{(1)}$ & $d_{dd}^{(1)}$ & $d_{ud}^{(1)}$ & $d_{ud}^{(8)}$ &
$d_{qq}^{(1)}$ & $d_{qq}^{(8)}$ & $d_{qu}^{(1)}$ & $d_{qu}^{(8)}$ & $d_{qd}^{(1)}$ & $d_{qd}^{(8)}$ \\
\hline
all models & $-\frac{1}{6}$ & $-\frac{1}{6}$ & $0$ & $-1$ & $0$ & $-\frac{1}{2}$ & $0$ & $-1$ & $0$ & $-1$\\
\hline
\end{tabular}
\caption{Coefficients $c_{ab}^{(1,8)}$ relevant for dijet bounds in the doublet, triplet and bidoublet models as well as the coefficients $d_{ab}^{(1,8)}$, which are independent of the flavour and electroweak groups.}
\label{tab:dijet}
\end{table}

\section{Production and decay of vector resonances}\label{sec:app-dijet}

The production cross section of a gluon resonance in $pp$ collisions reads
\begin{align}
\sigma(pp\to G^*) &=  \frac{2\pi}{9s}
\left[
(|g_L^u|^2+|g_R^u|^2) \mathcal L_{u\bar u}(s,m_\rho^2)
+
(|g_L^d|^2+|g_R^d|^2) \mathcal L_{d\bar d}(s,m_\rho^2)
\right],
\label{eq:sigmapprho}
\end{align}
where
\begin{equation}
\mathcal L_{q\bar q}(s,\hat s)=
\int_{\hat s/s}^1 \frac{dx}{x}
f_q(x,\mu)
f_{\bar q}\!\left(\frac{\hat s}{xs},\mu\right)
\end{equation}
is the parton-parton luminosity function at partonic (hadronic) center of mass energy $\sqrt{\hat s}$ ($\sqrt{s}$)
and the couplings $g_{L,R}^{u,d}$ are defined as $\mathcal L \supset \bar u_L \gamma^\mu T^a g_L^u
u_L G^*_\mu$ and can be read off eq.~\ref{eq:gG}. Again, there is a contribution due to fermion mixing, which is only relevant in $U(3)^3$ models due to the potentially sizable compositeness of the first generation, while the contribution due to vector mixing is always present.
The total width reads
\begin{equation}
\Gamma(G^*\to q \bar q) = \sum_{q=u,d}\sum_{i=1}^3\frac{m_\rho}{48\pi}\left(|g^{q^i}_L|^2+|g^{q^i}_R|^2\right)
\end{equation}
while the branching ratio to dijets is simply the width without the top contribution divided by the total width\footnote{Neglecting the top quark mass in the kinematics, which is a good approximation for multi-TeV resonances still allowed by the constraints}.

\section{Chirality-conserving flavour-changing interactions in $U(3)^3$}
\label{sec:app-U2U3}

\newcommand{\ELqL}{\boldsymbol{q}_{\boldsymbol{L}}}
\newcommand{\ELuR}{\boldsymbol{u}_{\boldsymbol{R}}}
\newcommand{\ELdR}{\boldsymbol{d}_{\boldsymbol{R}}}
\newcommand{\EHqL}{q_{3L}}
\newcommand{\EHuR}{t_{R}}
\newcommand{\EHdR}{b_{R}}
\newcommand{\ELqLbar}{\boldsymbol{\bar q}_{\boldsymbol{L}}}
\newcommand{\ELuRbar}{\boldsymbol{\bar u}_{\boldsymbol{R}}}
\newcommand{\ELdRbar}{\boldsymbol{\bar d}_{\boldsymbol{R}}}
\newcommand{\EHqLbar}{\bar q_{3L}}
\newcommand{\EHuRbar}{\bar t_{R}}
\newcommand{\EHdRbar}{\bar b_{R}}
\newcommand{\CLquL}{\boldsymbol{Q_L^u}}
\newcommand{\CLqdL}{\boldsymbol{Q_L^d}}
\newcommand{\CLquR}{\boldsymbol{Q_R^u}}
\newcommand{\CLqdR}{\boldsymbol{Q_R^d}}
\newcommand{\CLuL}{\boldsymbol{U_L}}
\newcommand{\CLdL}{\boldsymbol{D_L}}
\newcommand{\CHquL}{Q_{3L}^u}
\newcommand{\CHqdL}{Q_{3L}^d}
\newcommand{\CHquR}{Q_{3R}^u}
\newcommand{\CHqdR}{Q_{3R}^d}
\newcommand{\CHuR}{T_{R}}
\newcommand{\CHdR}{B_{R}}
\newcommand{\CLquLbar}{\boldsymbol{\bar Q_L^u}}
\newcommand{\CLqdLbar}{\boldsymbol{\bar Q_L^d}}
\newcommand{\CLquRbar}{\boldsymbol{\bar Q_R^u}}
\newcommand{\CLqdRbar}{\boldsymbol{\bar Q_R^d}}
\newcommand{\CLuLbar}{\boldsymbol{\bar U_L}}
\newcommand{\CLdLbar}{\boldsymbol{\bar D_L}}
\newcommand{\CHquLbar}{\bar Q_{3L}^u}
\newcommand{\CHqdLbar}{\bar Q_{3L}^d}
\newcommand{\CHuLbar}{\bar T_{L}}
\newcommand{\CHdLbar}{\bar B_{L}}
\newcommand{\V}{\boldsymbol{V}}

In $U(3)^3_\text{RC}$  the effective Yukawa couplings have the form 
\begin{equation}
\bar{q}_L \hat{s}_{Lu} Y_U s_{Ru} u_R
\label{Yuk_RC}
\end{equation}
 (and similarly for the down quarks) where $\hat{s}_{Lu}$ is a generic $3\times 3$ mixing matrix and $Y_U$, $s_{Ru}$ are both proportional to the unit matrix. In $U(3)^3_\text{LC}$ the role of the mixings is reversed and the Yukawa couplings take the form  
\begin{equation}
\bar{q}_L {s}_{Lu} Y_U \hat{s}_{Ru} u_R.
\label{Yuk_LC}
\end{equation}
At the same time, before going to the physical basis, the relevant interactions with the composite resonances have the form in $U(3)^3_\text{RC}$ 
\begin{equation}
\rho_\mu (\bar{q}_L \hat{s}_{Lu} \gamma_\mu \hat{s}_{Lu}^{\dag} q_L)
\label{int_RC}
\end{equation}
and in $U(3)^3_\text{LC}$ 
\begin{equation}
\rho_\mu (\bar{q}_L {s}_{Lu} \gamma_\mu {s}_{Lu}^* q_L).
\label{int_LC}
\end{equation}
In $U(3)^3_\text{RC}$ the physical bases for up and down quarks are reached by proper $3\times 3$ unitary transformations that diagonalize $\hat{s}_{Lu}$ and $\hat{s}_{Ld}$
\begin{equation}
U_L^u\hat{s}_{Lu} U_R^{u\dag}=  \hat{s}_{Lu}^{\rm diag}~~~~ U_L^d\hat{s}_{Ld} U_R^{d\dag}=  \hat{s}_{Ld}^{\rm diag},
\end{equation} 
 so that the CKM matrix is $V = U_L^u U_L^{d\dag}$. In the same physical basis the interaction  (\ref{int_RC}) in the down sector becomes
\begin{equation}
\rho_\mu (\bar{d}_L V^{\dag} \hat{s}_{Lu}^{\rm diag} \gamma_\mu (\hat{s}_{Lu}^{\rm diag})^* V d_L) \approx 
\rho_\mu s_{Lt}^2 \xi_{ij} (\bar{d}_{Li}  \gamma_\mu  d_{Lj}),~~~~~\xi_{ij} = V_{ti}^* V_{tj},
\label{int_dRC}
\end{equation}
which explains (\ref{U(3)RC-FC}). Note that the ratio of the third to  the second entry in  $\hat{s}_{Lu}^{\rm diag}$ equals $y_t/y_c$. On the other hand a similar procedure for $U(3)^3_\text{LC}$ leaves (\ref{int_LC}) unaltered since ${s}_{Lu}$ is proportional to the identity matrix.

\section{$U(2)^3$ in composite models}\label{sec:app-U2}
For ease of the reader we recall the setup of $U(2)^3$.
The strong sector can be taken invariant under a $U(2)_{Q+U+D}$ flavour symmetry acting on the first two generations of composite quarks. In right-compositeness -- meaningful only in the bidoublet model -- in order to generate the CKM matrix one has to consider a larger $U(2)_{Q^u+U}\times U(2)_{Q^d+D}$ symmetry.
Let us define
\begin{align}
Q ^{u} &= \begin{pmatrix}\boldsymbol{Q^{u}}\\ Q_3^{u}\end{pmatrix}, & U &= \begin{pmatrix}\boldsymbol{U}\\T \end{pmatrix}, & q_L &= \begin{pmatrix}\ELqL\\ \EHqL\end{pmatrix}, & u_R &= \begin{pmatrix}\ELuR\\ \EHuR\end{pmatrix},
\end{align}
where the first two generation doublets are written in boldface, and the same for down-type quarks. The mixing Lagrangians in the cases of {\it left-compositeness} and {\it right-compositeness} are respectively\footnote{We write the Lagrangians for the bidoublet model. The doublet and triplet cases are analogous, with $Q^u$ and $Q^d$ replaced by a single $Q$.}
\begin{align}
\mathcal{L}_\text{mix}^{U(2)^3_\text{LC}} &=
m_{U3}\lambda_{Lu3}\, \EHqLbar \CHquR +
m_{U2}\lambda_{Lu2} \,\ELqLbar\CLquR +
m_{U3}\lambda_{Ru3}\, \CHuLbar \EHuR
\notag\\
&+m_{U2}\,d_u\, (\CLuLbar\V)\EHuR +
m_{U2}\,\CLuLbar \Delta_u \ELuR +
\text{h.c.}
+ (u,U,t,T\to d,D,b,B)
\label{mixing2L}
\end{align}
and
\begin{align}
\mathcal{L}_\text{mix}^{U(2)^3_\text{RC}} &=
m_{U3}\lambda_{Ru3}\, \CHuLbar\EHuR +
m_{U2}\lambda_{Ru2} \, \CLuLbar\ELuR +
m_{U3}\lambda_{L(u)3}\, \EHqLbar\CHquR
\notag\\
&+m_{U3}\,d_u\, (\ELqLbar\V)\CHquR +
m_{U2}\,\ELqLbar \Delta_u \CLquR +
\text{h.c.}
+ (u,U,t,T\to d,D,b,B).
\label{mixing2R}
\end{align}
The mixings in the first line of \eqref{mixing2L} and \eqref{mixing2R} break the symmetry of the strong sector down to $U(2)_q\times U(2)_u\times U(2)_d$. This symmetry is in turn broken minimally by the spurions
\begin{align}
\V&\sim({\bf 2},{\bf 1},{\bf 1}),& \Delta_u&\sim({\bf 2},{\bf 2},{\bf 1}),& \Delta_d&\sim({\bf 2},{\bf 1},{\bf 2}).
\label{MU2spurions}
\end{align}
Using $U(2)^3$ transformations of the quarks they can be put in the simple form
\begin{align}
\V &= \begin{pmatrix} 0 \\ \epsilon_L \end{pmatrix},&
\Delta_u &=
\begin{pmatrix}
c_u & s_u e^{i\alpha_u} \\
-s_u e^{-i\alpha_u} & c_u 
\end{pmatrix}
\begin{pmatrix}
\lambda_{Xu1} & 0 \\
0 & \lambda_{Xu2}
\end{pmatrix}, & (u\leftrightarrow d),
\end{align}
where $X=R,L$ in left- and right-compositeness, respectively. Here we do not discuss the case of generic $U(2)^3$ breaking introduced in \cite{Barbieri:2012bh}.

The SM Yukawa couplings \eqref{SMYuk} can be written in terms of the spurions as
\begin{align}\label{SMYukLC}
\hat y_u &= \begin{pmatrix}a_u\, \Delta_u & b_t e^{i\phi_t}\V\\ 0 & y_t\end{pmatrix}, &
\hat y_d &= \begin{pmatrix}a_d\, \Delta_d & b_b e^{i\phi_b}\V\\ 0 & y_b\end{pmatrix},
\end{align}
where
\begin{align}
y_t &= Y_{U3}s_{Lu3}s_{Ru3},
\\
a_u &= Y_{U2}s_{Lu2},& b_t &= Y_{U2}s_{Lu2}\,d_u, & &\text{ in left-compositeness},\\ a_u &= Y_{U2}s_{Ru2},& b_t &= Y_{U3}s_{Ru3}\,d_u,& &\text{ in right-compositeness},
\end{align}
$s_{Xi} = \lambda_{Xi}/\sqrt{1 + (\lambda_{Xi})^2}$, and similarly for $a_d$, $b_b$ and $y_b$.
Here and in the following we consider all the parameters real, factoring out the phases everywhere as in \eqref{SMYukLC}. The $\hat y_{u,d}$ are diagonalized to a sufficient level of approximation by pure unitary transformations of the left-handed quarks \cite{Barbieri:2012uh}
\begin{align}\label{UL}
U_u &\simeq \begin{pmatrix}c_{u} & s_u e^{i\alpha_u} & -s_u s_t e^{i(\alpha_u + \phi_t)}\\
-s_u e^{-i\alpha_u} & c_u & -c_u s_t e^{i\phi_t}\\
0 & s_t e^{-i\phi_t} & 1
\end{pmatrix}, &
U_d &\simeq \begin{pmatrix}c_{d} & s_d e^{i\alpha_d} & -s_d s_b e^{i(\alpha_d + \phi_b)}\\
-s_d e^{-i\alpha_d} & c_d & -c_d s_b e^{i\phi_b}\\
0 & s_b e^{-i\phi_b} & 1
\end{pmatrix},\end{align}
where
\begin{align}
s_t &= Y_{U2}s_{Lu2}\frac{d_u\epsilon_L}{y_t}, & s_b &= Y_{D2}s_{Ld2}\frac{d_d\epsilon_L}{y_b},& &\text{in left-compositeness},
\label{eq:stbLC}
\\
s_t &= Y_{U3}s_{Ru3}\frac{d_u\epsilon_L}{y_t}, & s_b &= Y_{D3}s_{Rd3}\frac{d_d\epsilon_L}{y_b},& &\text{in right-compositeness}.
\end{align}

The CKM matrix is $V = U_uU_d^{\dag}$ and, after a suitable redefinition of quark phases, takes the form
\begin{equation}
V  \simeq
\left(\begin{array}{ccc}
 1- \lambda^2/2 &  \lambda & s_u s e^{-i \delta}  \\
-\lambda & 1- \lambda^2/2   & c_u s  \\
-s_d s \,e^{i (\phi+\delta)} & -s c_d & 1 \\
\end{array}\right),
\label{eq:CKMstand}
\end{equation}
where
\begin{align}
s_uc_d - c_us_d e^{-i\phi} &\equiv \lambda e^{i\delta}, & s_b e^{i\phi_b} - s_t e^{i\phi_t}\equiv s e^{i\chi}.
\end{align}

\subsection*{Chirality-conserving flavour-changing interactions}
Equations (\ref{int_RC}, \ref{int_LC}) remain formally true in $U(2)^3$  as well, with the following qualifications.  $Y_U, s_{Ru},  s_{Lu}$ are no longer proportional to the identity but are still  diagonal with only the first  two entries equal to each other. At the same time minimal breaking of $U(2)^3$ leads to a special form of the matrices $ \hat{s}_{Lu},  \hat{s}_{Ru}$ that allows to diagonalize approximately the Yukawa couplings  by pure left unitary transformations of the form \eqref{UL}.

In $U(2)^3_\text{RC}$ these transformations lead to exactly the same equation as (\ref{int_dRC}), whereas in the $U(2)^3_\text{LC}$ case equation (\ref{int_LC}) in the down sector goes into
\begin{equation}
\rho_\mu (\bar{d}_L U_d {s}_{Lu} \gamma_\mu {s}_{Lu}^* U_{d}^{\dag} d_L) \approx 
\rho_\mu s_{Lt}^2 \chi_{ij} (\bar{d}_{Li}  \gamma_\mu  d_{Lj}),~~~~~\chi_{ij} = U^d_{i3} U^{d*}_{j3},
\label{int_dLC}
\end{equation}
Remember that, contrary to the $U(3)^3_\text{RC}$ case, ${s}_{Lu}$, although still diagonal, is not proportional to the unit matrix. Hence a flavour violation survives as in (\ref{FCU2LC}) with
\begin{equation}
\label{eq:rb}
r_b = \frac{s_b}{s} e^{i(\chi - \phi_b)}.
\end{equation}

Note that in \UdueLC, for $Y_{U2}\sim Y_{D2}\sim O(1)$ and $d_u,d_d \lesssim O(1)$, (\ref{eq:stbLC}) leads to two possibilities:
\begin{enumerate}
 \item $s_t\ll s_b$, i.e. $|r_b|\approx1$;
 \item $s_t\sim s_b\sim|V_{cb}|$, which allows $|r_b|$ to deviate from 1 but requires at the same time $s_{Lu2}\epsilon_L\sim |V_{cb}|$.
\end{enumerate}
In the first case one would have $m_{\psi} \gtrsim 1\text{--}1.5$ TeV from the flavour bounds of table \ref{tab:bounds-U2LC}, while in the second case one can obtain a minimal value of $m_\psi \simeq 0.6$ TeV, as in table \ref{tab:mmin}, for $|r_b| \sim 0.25$ and $Y \sim 1$. However, to avoid a too large $U(2)^3$-breaking -- i.e. a large $\epsilon_L$ -- the mixing angle of the first generations quarks $s_{Lu2}$ cannot be too small. This in turn has to be confronted with the lower bounds on $m_\psi$ from $R_h$, $V_{\text{CKM}}$ and the dijet angular distribution shown in table \ref{tab:bounds-U2LC-sL2}: to make them consistent with $m_\psi \simeq 0.6$ TeV, it must 
be $\epsilon_L \gtrsim 0.3$. Note anyhow that we are not treating $\epsilon_L$ as an expansion parameter.
\begin{table}[tbp]
\renewcommand{\arraystretch}{1.3}
\centering
\begin{tabular}{cccc}
\hline
Observable & \multicolumn{3}{c}{Bounds on $m_{\psi}$ [TeV]} \\
&  doublet & triplet  & bidoublet\\
\hline
$R_h$ & $7.2 ~s_{L2}Y_2$  &$6.8 ~s_{L2}Y_2$  &$5.6 ~s_{Lu2}Y_{U2}$\\
$V_\text{CKM}$ & $8.4 ~s_{L2}Y_2$  &$6.8 ~s_{L2}Y_2$  &$6.8 ~s_{Lu2}Y_{U2}$\\
$pp\to jj$ ang. dist. & $4.3  ~s_{L2}^2Y_{2}$ & $5.3  ~s_{L2}^2Y_{2}$ & $5.3  ~s_{Lu2}^2Y_{U2}$  \\
\hline
\end{tabular}
\caption{Lower bounds on the fermion resonance mass $m_\psi$ in TeV in \UdueLC\ from left-handed 1st and 2nd generation quarks mixed with the composite resonances by an angle $s_{Lu2}$.}
\label{tab:bounds-U2LC-sL2}
\end{table}

In the bidoublet model, in addition to \eqref{int_dLC} there are also the terms coming from the mixing with the $\bar Q^d\gamma_{\mu}Q^d$ current, which are suppressed as $1/z^2$.
In the up-quark sector with right-compositeness only this suppressed contribution from $Q^d$ gives rise to flavour violation, while in left-compositeness the analog of \eqref{int_dLC} holds, with $U_d$ replaced by $U_u$.

Flavour violation from chirality-conserving right-handed quark bilinears is instead suppressed, a general property of the Minimal $U(2)^3$ framework \cite{Barbieri:2011ci,Barbieri:2012uh}.

\bibliographystyle{My}
\bibliography{chm}

\providecommand{\href}[2]{#2}\begingroup\raggedright\begin{thebibliography}{10}

\bibitem{:2012gk}
{\bf ATLAS Collaboration}, G.~Aad {\em et al.,}
  \href{http://dx.doi.org/10.1016/j.physletb.2012.08.020}{{\em Phys.Lett.} {\bf
  B716} (2012)  1--29},
\href{http://arxiv.org/abs/1207.7214}{{\tt arXiv:1207.7214 [hep-ex]}}.

\bibitem{:2012gu}
{\bf CMS Collaboration}, S.~Chatrchyan {\em et al.,}
  \href{http://dx.doi.org/10.1016/j.physletb.2012.08.021}{{\em Phys.Lett.} {\bf
  B716} (2012)  30--61},
\href{http://arxiv.org/abs/1207.7235}{{\tt arXiv:1207.7235 [hep-ex]}}.

\bibitem{Kaplan:1983fs}
D.~B. Kaplan and H.~Georgi,
\href{http://dx.doi.org/10.1016/0370-2693(84)91177-8}{{\em Phys.Lett.} {\bf
  B136} (1984)  183}.

\bibitem{Georgi:1984af}
H.~Georgi and D.~B. Kaplan,
\href{http://dx.doi.org/10.1016/0370-2693(84)90341-1}{{\em Phys.Lett.} {\bf
  B145} (1984)  216}.

\bibitem{Contino:2003ve}
R.~Contino, Y.~Nomura, and A.~Pomarol,
  \href{http://dx.doi.org/10.1016/j.nuclphysb.2003.08.027}{{\em Nucl.Phys.}
  {\bf B671} (2003)  148--174},
\href{http://arxiv.org/abs/hep-ph/0306259}{{\tt arXiv:hep-ph/0306259
  [hep-ph]}}.

\bibitem{Agashe:2004rs}
K.~Agashe, R.~Contino, and A.~Pomarol,
  \href{http://dx.doi.org/10.1016/j.nuclphysb.2005.04.035}{{\em Nucl.Phys.}
  {\bf B719} (2005)  165--187},
\href{http://arxiv.org/abs/hep-ph/0412089}{{\tt arXiv:hep-ph/0412089
  [hep-ph]}}.

\bibitem{Kaplan:1991dc}
D.~B. Kaplan
\href{http://dx.doi.org/10.1016/S0550-3213(05)80021-5}{{\em Nucl.Phys.} {\bf
  B365} (1991)  259--278}.

\bibitem{Contino:2006qr}
R.~Contino, L.~{Da Rold}, and A.~Pomarol,
  \href{http://dx.doi.org/10.1103/PhysRevD.75.055014}{{\em Phys.Rev.} {\bf D75}
  (2007)  055014},
\href{http://arxiv.org/abs/hep-ph/0612048}{{\tt arXiv:hep-ph/0612048
  [hep-ph]}}.

\bibitem{Pomarol:2012qf}
A.~Pomarol and F.~Riva, \href{http://dx.doi.org/10.1007/JHEP08(2012)135}{{\em
  JHEP} {\bf 1208} (2012)  135},
\href{http://arxiv.org/abs/1205.6434}{{\tt arXiv:1205.6434 [hep-ph]}}.

\bibitem{Redi:2012ha}
M.~Redi and A.~Tesi, \href{http://dx.doi.org/10.1007/JHEP10(2012)166}{{\em
  JHEP} {\bf 1210} (2012)  166},
\href{http://arxiv.org/abs/1205.0232}{{\tt arXiv:1205.0232 [hep-ph]}}.

\bibitem{Matsedonskyi:2012ym}
O.~Matsedonskyi, G.~Panico, and A.~Wulzer,
\href{http://arxiv.org/abs/1204.6333}{{\tt arXiv:1204.6333 [hep-ph]}}.

\bibitem{Marzocca:2012zn}
D.~Marzocca, M.~Serone, and J.~Shu,
  \href{http://dx.doi.org/10.1007/JHEP08(2012)013}{{\em JHEP} {\bf 1208} (2012)
   013},
\href{http://arxiv.org/abs/1205.0770}{{\tt arXiv:1205.0770 [hep-ph]}}.

\bibitem{Panico:2012uw}
G.~Panico, M.~Redi, A.~Tesi, and A.~Wulzer,
\href{http://arxiv.org/abs/1210.7114}{{\tt arXiv:1210.7114 [hep-ph]}}.

\bibitem{CMS:2012ab}
{\bf CMS Collaboration}, S.~Chatrchyan {\em et al.,}
  \href{http://dx.doi.org/10.1016/j.physletb.2012.07.059}{{\em Phys.Lett.} {\bf
  B716} (2012)  103--121},
\href{http://arxiv.org/abs/1203.5410}{{\tt arXiv:1203.5410 [hep-ex]}}.

\bibitem{Chatrchyan:2012vu}
{\bf CMS Collaboration}, S.~Chatrchyan {\em et al.,}
\href{http://arxiv.org/abs/1209.0471}{{\tt arXiv:1209.0471 [hep-ex]}}.

\bibitem{Chatrchyan:2012af}
{\bf CMS Collaboration}, S.~Chatrchyan {\em et al.,}
\href{http://arxiv.org/abs/1210.7471}{{\tt arXiv:1210.7471 [hep-ex]}}.

\bibitem{ATLAS-CONF-2012-130}
{\bf ATLAS Collaboration} Tech. Rep. ATLAS-CONF-2012-130, CERN, Geneva, Sep,
  2012.

\bibitem{Contino:2006nn}
R.~Contino, T.~Kramer, M.~Son, and R.~Sundrum,
  \href{http://dx.doi.org/10.1088/1126-6708/2007/05/074}{{\em JHEP} {\bf 0705}
  (2007)  074},
\href{http://arxiv.org/abs/hep-ph/0612180}{{\tt arXiv:hep-ph/0612180
  [hep-ph]}}.

\bibitem{Grossman:1999ra}
Y.~Grossman and M.~Neubert,
  \href{http://dx.doi.org/10.1016/S0370-2693(00)00054-X}{{\em Phys.Lett.} {\bf
  B474} (2000)  361--371},
\href{http://arxiv.org/abs/hep-ph/9912408}{{\tt arXiv:hep-ph/9912408
  [hep-ph]}}.

\bibitem{Huber:2000ie}
S.~J. Huber and Q.~Shafi,
  \href{http://dx.doi.org/10.1016/S0370-2693(00)01399-X}{{\em Phys.Lett.} {\bf
  B498} (2001)  256--262},
\href{http://arxiv.org/abs/hep-ph/0010195}{{\tt arXiv:hep-ph/0010195
  [hep-ph]}}.

\bibitem{Gherghetta:2000qt}
T.~Gherghetta and A.~Pomarol,
  \href{http://dx.doi.org/10.1016/S0550-3213(00)00392-8}{{\em Nucl.Phys.} {\bf
  B586} (2000)  141--162},
\href{http://arxiv.org/abs/hep-ph/0003129}{{\tt arXiv:hep-ph/0003129
  [hep-ph]}}.

\bibitem{Agashe:2004cp}
K.~Agashe, G.~Perez, and A.~Soni,
  \href{http://dx.doi.org/10.1103/PhysRevD.71.016002}{{\em Phys.Rev.} {\bf D71}
  (2005)  016002},
\href{http://arxiv.org/abs/hep-ph/0408134}{{\tt arXiv:hep-ph/0408134
  [hep-ph]}}.

\bibitem{Blanke:2008zb}
M.~Blanke, A.~J. Buras, B.~Duling, S.~Gori, and A.~Weiler,
  \href{http://dx.doi.org/10.1088/1126-6708/2009/03/001}{{\em JHEP} {\bf 0903}
  (2009)  001},
\href{http://arxiv.org/abs/0809.1073}{{\tt arXiv:0809.1073 [hep-ph]}}.

\bibitem{Bauer:2009cf}
M.~Bauer, S.~Casagrande, U.~Haisch, and M.~Neubert,
  \href{http://dx.doi.org/10.1007/JHEP09(2010)017}{{\em JHEP} {\bf 1009} (2010)
   017},
\href{http://arxiv.org/abs/0912.1625}{{\tt arXiv:0912.1625 [hep-ph]}}.

\bibitem{KerenZur:2012fr}
B.~Keren-Zur, P.~Lodone, M.~Nardecchia, D.~Pappadopulo, R.~Rattazzi, {\em et
  al.,} \href{http://dx.doi.org/10.1016/j.nuclphysb.2012.10.012}{{\em
  Nucl.Phys.} {\bf B867} (2013)  429--447},
\href{http://arxiv.org/abs/1205.5803}{{\tt arXiv:1205.5803 [hep-ph]}}.

\bibitem{Csaki:2008zd}
C.~Csaki, A.~Falkowski, and A.~Weiler,
  \href{http://dx.doi.org/10.1088/1126-6708/2008/09/008}{{\em JHEP} {\bf 0809}
  (2008)  008},
\href{http://arxiv.org/abs/0804.1954}{{\tt arXiv:0804.1954 [hep-ph]}}.

\bibitem{Cacciapaglia:2007fw}
G.~Cacciapaglia, C.~Csaki, J.~Galloway, G.~Marandella, J.~Terning, {\em et
  al.,} \href{http://dx.doi.org/10.1088/1126-6708/2008/04/006}{{\em JHEP} {\bf
  0804} (2008)  006},
\href{http://arxiv.org/abs/0709.1714}{{\tt arXiv:0709.1714 [hep-ph]}}.

\bibitem{Barbieri:2008zt}
R.~Barbieri, G.~Isidori, and D.~Pappadopulo,
  \href{http://dx.doi.org/10.1088/1126-6708/2009/02/029}{{\em JHEP} {\bf 0902}
  (2009)  029},
\href{http://arxiv.org/abs/0811.2888}{{\tt arXiv:0811.2888 [hep-ph]}}.

\bibitem{Redi:2011zi}
M.~Redi and A.~Weiler, \href{http://dx.doi.org/10.1007/JHEP11(2011)108}{{\em
  JHEP} {\bf 1111} (2011)  108},
\href{http://arxiv.org/abs/1106.6357}{{\tt arXiv:1106.6357 [hep-ph]}}.

\bibitem{Barbieri:2011ci}
R.~Barbieri, G.~Isidori, J.~Jones-Perez, P.~Lodone, and D.~M. Straub,
  \href{http://dx.doi.org/10.1140/epjc/s10052-011-1725-z}{{\em Eur.Phys.J.}
  {\bf C71} (2011)  1725},
\href{http://arxiv.org/abs/1105.2296}{{\tt arXiv:1105.2296 [hep-ph]}}.

\bibitem{Barbieri:2012uh}
R.~Barbieri, D.~Buttazzo, F.~Sala, and D.~M. Straub,
  \href{http://dx.doi.org/10.1007/JHEP07(2012)181}{{\em JHEP} {\bf 1207} (2012)
   181},
\href{http://arxiv.org/abs/1203.4218}{{\tt arXiv:1203.4218 [hep-ph]}}.

\bibitem{Barbieri:2007bh}
R.~Barbieri, B.~Bellazzini, V.~S. Rychkov, and A.~Varagnolo,
  \href{http://dx.doi.org/10.1103/PhysRevD.76.115008}{{\em Phys.Rev.} {\bf D76}
  (2007)  115008},
\href{http://arxiv.org/abs/0706.0432}{{\tt arXiv:0706.0432 [hep-ph]}}.

\bibitem{Baak:2012kk}
M.~Baak, M.~Goebel, J.~Haller, A.~Hoecker, D.~Kennedy, {\em et al.,}
\href{http://arxiv.org/abs/1209.2716}{{\tt arXiv:1209.2716 [hep-ph]}}.

\bibitem{Agashe:2006at}
K.~Agashe, R.~Contino, L.~{Da Rold}, and A.~Pomarol,
  \href{http://dx.doi.org/10.1016/j.physletb.2006.08.005}{{\em Phys.Lett.} {\bf
  B641} (2006)  62--66},
\href{http://arxiv.org/abs/hep-ph/0605341}{{\tt arXiv:hep-ph/0605341
  [hep-ph]}}.

\bibitem{Vignaroli:2012si}
N.~Vignaroli
\href{http://arxiv.org/abs/1204.0478}{{\tt arXiv:1204.0478 [hep-ph]}}.

\bibitem{Altmannshofer:2012az}
W.~Altmannshofer and D.~M. Straub,
  \href{http://dx.doi.org/10.1007/JHEP08(2012)121}{{\em JHEP} {\bf 1208} (2012)
   121},
\href{http://arxiv.org/abs/1206.0273}{{\tt arXiv:1206.0273 [hep-ph]}}.

\bibitem{Isidori:2011qw}
G.~Isidori, J.~F. Kamenik, Z.~Ligeti, and G.~Perez,
  \href{http://dx.doi.org/10.1016/j.physletb.2012.03.046}{{\em Phys.Lett.} {\bf
  B711} (2012)  46--51},
\href{http://arxiv.org/abs/1111.4987}{{\tt arXiv:1111.4987 [hep-ph]}}.

\bibitem{Calibbi:2012at}
L.~Calibbi, Z.~Lalak, S.~Pokorski, and R.~Ziegler,
  \href{http://dx.doi.org/10.1007/JHEP07(2012)004}{{\em JHEP} {\bf 1207} (2012)
   004},
\href{http://arxiv.org/abs/1204.1275}{{\tt arXiv:1204.1275 [hep-ph]}}.

\bibitem{Buras:2011ph}
A.~J. Buras, C.~Grojean, S.~Pokorski, and R.~Ziegler,
  \href{http://dx.doi.org/10.1007/JHEP08(2011)028}{{\em JHEP} {\bf 1108} (2011)
   028},
\href{http://arxiv.org/abs/1105.3725}{{\tt arXiv:1105.3725 [hep-ph]}}.

\bibitem{Agashe:2008uz}
K.~Agashe, A.~Azatov, and L.~Zhu,
  \href{http://dx.doi.org/10.1103/PhysRevD.79.056006}{{\em Phys.Rev.} {\bf D79}
  (2009)  056006},
\href{http://arxiv.org/abs/0810.1016}{{\tt arXiv:0810.1016 [hep-ph]}}.

\bibitem{Gedalia:2009ws}
O.~Gedalia, G.~Isidori, and G.~Perez,
  \href{http://dx.doi.org/10.1016/j.physletb.2009.10.097}{{\em Phys.Lett.} {\bf
  B682} (2009)  200--206},
\href{http://arxiv.org/abs/0905.3264}{{\tt arXiv:0905.3264 [hep-ph]}}.

\bibitem{Barbieri:2012bh}
R.~Barbieri, D.~Buttazzo, F.~Sala, and D.~M. Straub,
  \href{http://dx.doi.org/10.1007/JHEP10(2012)040}{{\em JHEP} {\bf 1210} (2012)
   040},
\href{http://arxiv.org/abs/1206.1327}{{\tt arXiv:1206.1327 [hep-ph]}}.

\bibitem{ATLAS-CONF-2012-088}
 Tech. Rep. ATLAS-CONF-2012-088, CERN, Geneva, Jul, 2012.

\bibitem{CMS-PAS-EXO-12-016}
 Tech. Rep. CMS-PAS-EXO-12-016, 2012.

\bibitem{Redi:2012uj}
M.~Redi \href{http://dx.doi.org/10.1140/epjc/s10052-012-2030-1}{{\em
  Eur.Phys.J.} {\bf C72} (2012)  2030},
\href{http://arxiv.org/abs/1203.4220}{{\tt arXiv:1203.4220 [hep-ph]}}.

\bibitem{Domenech:2012ai}
O.~Domenech, A.~Pomarol, and J.~Serra,
  \href{http://dx.doi.org/10.1103/PhysRevD.85.074030}{{\em Phys.Rev.} {\bf D85}
  (2012)  074030},
\href{http://arxiv.org/abs/1201.6510}{{\tt arXiv:1201.6510 [hep-ph]}}.

\bibitem{Altmannshofer:2008dz}
W.~Altmannshofer, P.~Ball, A.~Bharucha, A.~J. Buras, D.~M. Straub, {\em et
  al.,} \href{http://dx.doi.org/10.1088/1126-6708/2009/01/019}{{\em JHEP} {\bf
  0901} (2009)  019},
\href{http://arxiv.org/abs/0811.1214}{{\tt arXiv:0811.1214 [hep-ph]}}.

\end{thebibliography}\endgroup

\end{document}